\documentclass[aps, twocolumn, superscriptaddress, floatfix, longbibliography]{revtex4-1}
\usepackage[charter]{mathdesign}
\usepackage{amsmath,graphicx,bm}
\usepackage{tikz}
\usepackage[caption=false]{subfig}
\usepackage[pdftex,plainpages=false,colorlinks=true,linkcolor=blue, citecolor=blue, urlcolor=blue]{hyperref}
\newcommand{\etal}{\textit{et al.}}

\newcommand{\crs}[1][]{c_{ \textbf{\scriptsize r}#1\scriptsize,\sigma}}

\newcommand\av{\mathbf{a}}

\newcommand\rv{\mathbf{r}}
\newcommand\tv{\mathbf{t}}

\newcommand\qv{\mathbf{q}}
\newcommand\kvt{\mathbf{\tilde k}}

\newcommand\xh{\mathbf{\hat x}}
\newcommand\yh{\mathbf{\hat y}}

\newcommand\Gv{\mathbf{G}}

\DeclareMathOperator{\Tr}{Tr}

\newcommand\Sigmav{\bm{\Sigma}}

\newcommand\s{\sigma}

\newcommand\Gammav{\bm{\Gamma}}
\newcommand\thetav{\bm{\theta}}
\newcommand\epsilonv{\bm{\epsilon}}

\newcommand\omv{\bm{\omega}}

\begin{document}
\title{Charge- and pair-density-wave orders in the one-band Hubbard model with dynamical mean field theory}

\author{S. S. Dash}\affiliation{D\'epartement de physique and Institut quantique, Universit\'e de Sherbrooke, Sherbrooke, Qu\'ebec, Canada J1K 2R1}
\author{D. S\'en\'echal}\affiliation{D\'epartement de physique and Institut quantique, Universit\'e de Sherbrooke, Sherbrooke, Qu\'ebec, Canada J1K 2R1} 
\date{\today}

\begin{abstract}
We study the charge-density-wave order and its competition with superconductivity in the one-band Hubbard model 
for high-$T_c$ superconducting cuprates. We use cluster dynamical mean field theory (CDMFT) at $T=0$.
The one-band Hubbard model only contains copper atoms, and a physical charge-density-wave with charge excesses located on the oxygen atoms manifests itself as a bond-density-wave (BDW) within this context. It arises purely out of local correlation effects and also leads to a $s'$-wave pair-density-wave in the presence of $d$-wave superconductivity.
The $d$-wave BDW is suppressed on increasing $U$ and is favored on increasing the magnitude of the second-neighbor hopping $t'$. 
Further, the $d$-wave BDW order is weakened when in competition with superconductivity, as seen in experiments. 
Additionally it behaves quite differently on varying $U$ in the superconducting state compared to the normal state.  
\end{abstract}

\maketitle

%================================================================================
\section{Introduction}

Charge-density-wave (CDW) order is ubiquitous among various families of high-$T_c$ superconducting cuprates 
(HTSC) \cite{ghiringhelli2012long,kohsaka2007intrinsic,fujita2014direct,comin2016resonant}. 
Although the exact role of this broken-symmetry state in high-$T_c$ superconductivity is 
not yet understood, it has been observed that the CDW order is in competition with 
superconductivity \cite{ghiringhelli2012long,huecker2014competing,chang2012direct,croft2014charge}. 
In the $T$-$p$ (temperature - hole doping) phase diagram, the CDW phase appears roughly between 
the antiferromagnetic insulator at very low doping and superconductivity at intermediate doping.

CDWs are observed to exist mostly as unidirectional domains \cite{kohsaka2007intrinsic,fujita2014direct,comin2016resonant} and are known to have a predominant $d$-wave form factor with weak $s$-wave and $s'$-wave components \cite{fujita2014direct,comin2015symmetry}. 
CDWs are generally incommensurate \cite{ghiringhelli2012long,comin2014charge,huecker2014competing,croft2014charge,chang2012direct,blackburn2013x} and the associated wave number decreases with hole doping \cite{huecker2014competing,blackburn2013x}. 
However, Kohsaka \etal\cite{kohsaka2007intrinsic} and Fujita \etal\cite{fujita2014direct} 
suggest that a locally commensurate CDW with wave vector $q=0.25$ accounts for the observations very well. 
Further, it was recently observed that a $s'$-wave Cooper pair-density-wave (PDW) order, with a wave vector $q=0.25$, exists in underdoped 
cuprates \cite{hamidian2016detection,ruan2018visualization}, arising from the 
coexistence of a $d$-wave CDW order and $d$-wave superconductivity \cite{hamidian2016detection}.

It has been widely considered that the CDW order is closely related to the pseudogap phenomenon in hole-doped cuprates \cite{fujita2014direct,atkinson2015charge,comin2016resonant,chang2012direct,comin2014charge}.
A supporting observation would be that the CDW wave vector $q$ connects electronic states near the antinodal 
regions where the effect of the pseudogap is strongest. 
Atkinson \etal\cite{atkinson2015charge} have shown, using a generalized random-phase approximation (RPA) on the three-band Hubbard model, 
that a CDW instability originates from a simple model of the Fermi surface in the pseudogap phase. 
However, many studies advocate against a direct causal relation between the CDW order and the pseudogap. 
For instance, H\"ucker \etal~\cite{huecker2014competing} and Croft \etal~\cite{croft2014charge} observed that 
the onset temperature for the CDW ($T_{\textrm{CDW}}$) is lower than the onset temperature for the pseudogap ($T^*$),
and also that $T^{*}$ increases monotonously with underdoping while $T_{\textrm{CDW}}$ decreases beyond an optimal value. 
Further, Badoux \etal~\cite{badoux2016change} observed that the CDW order starts at a doping value lower than the onset of the pseudogap. 
Verret \etal~\cite{verret2017subgap} concluded, from a mean-field study, that the CDWs alone cannot lead to the density of states seen in 
the pseudogap phase.

On the other hand, there has been some evidence that antiferromagnetic interactions are the key behind CDW order. 
For instance, Davis and Lee~\cite{davis2013concepts} start with an effective Hamiltonian with antiferromagnetic interactions and obtain the CDW instability along with other phases in the cuprate phase diagram.  
Various studies using the $t$-$J$ model and its variations have been able to obtain the CDW order~\cite{sau2014mean,sachdev2013bond,vojta2000competing,vojta2002superconducting,bejas2012possible,raczkowski2007unidirectional,allais2014density}. 
Other studies include a spin-fermion model with short-range magnetic interactions, which have been able to obtain CDWs~\cite{pepin2014pseudogap,wang2014charge,meier2014cascade}.  
Studies of the extended Hubbard model using dynamical mean field theory (DMFT) also indicate the existence of charge-ordered states \cite{pietig1999reentrant,tong2004charge,amaricci2010extended,allais2014auxiliary}. 
Cluster-based approaches applied on the Hubbard model have succeeded in obtaining a CDW order, including studies with the dynamical cluster approximation (DCA)~\cite{terletska2017charge} and the variational cluster approximation (VCA)~\cite{faye2017interplay}. 
Further, the PDW order has also been observed theoretically in presence of the CDW order and 
superconductivity~\cite{raczkowski2007unidirectional,faye2017interplay,freire2015renormalization}.

The antiferromagnetic origin of the CDW order should reflect in its dependence on the second-neighbor hopping $t'$, which causes magnetic frustration and is detrimental to antiferromagnetic fluctuations.
Vojta \etal~\cite{vojta2000competing} and White \etal~\cite{white1999competition} observe that a finite $t'$ weakens the charge-ordered state. 
However, Bejas \etal~\cite{bejas2012possible} observe the CDW order to strengthen as the magnitude of $t'$ increases up to a certain value, 
beyond which it starts to weaken. Also, Bauer $\&$ Hewson~\cite{bauer2010competition} observed a competition between the charge-ordered state and antiferromagnetism at half-filling in the Hubbard-Holstein model. 
Such conflicting observations indicate that the relation between antiferromagnetism and CDW order is not clear. 

The competition between CDW order and superconductivity has been observed as a function of temperature. 
At a doping value where $T_{\textrm{CDW}}> T_c$, the CDW intensity starts to grow at $T_{\textrm{CDW}}$ and increases upon lowering the temperature until $T_c$, below which it starts decreasing~\cite{ghiringhelli2012long,chang2012direct,croft2014charge,blackburn2013x,achkar2012distinct,blanco2013momentum}, indicating that it weakens as the superconducting (SC) order parameter grows. 
Further, when a magnetic field is applied below $T_c$, which weakens superconductivity, the CDW 
intensity grows with the magnetic field~\cite{huecker2014competing,chang2012direct,blackburn2013x,blanco2013momentum}. 
Various theoretical studies have also observed such a competition between the two phases~\cite{sau2014mean,pepin2014pseudogap,wang2014charge,meier2014cascade,faye2017interplay,cappelluti1999interplay}.

In this work, we study doped Mott insulators at zero temperature using cluster dynamical mean field theory (CDMFT) 
with an exact diagonalization (ED) impurity solver, applied to the one-band Hubbard model.
%\comment{Need to stress here the originality of our paper : dSC computed concurrently and dynamically, etc.}
%The Hubbard model is more realistic than the $t$-$J$ model in the intermediate-coupling range and describes charge fluctuations better.
We observe the $d$-wave SC order arising spontaneously out of the CDMFT self-consistency along with 
various density-wave (DW) orders ($s$-, $s'$- and $d$- wave), with a dominant $d$-wave 
bond-density-wave (BDW) order. 
We study these DW orders alone (normal state: no superconductivity) and in the SC state 
for various values of $U$ in the strong correlation regime. 
In the normal state, we find that the BDW order is suppressed on increasing $U$ and is favored by increasing the magnitude of $t'$. 
In the SC state, we observe that the BDW order is weakened compared to the normal state. 
Additionally, the SC order is also weakened in presence of the BDW order compared 
to superconductivity alone, indicating an inherent competition between the two orders as suggested by various experiments~\cite{chang2012direct,croft2014charge,ghiringhelli2012long}. 
Further, we also observe a PDW order when BDW and SC orders coexist as seen in experiments~\cite{hamidian2016detection,ruan2018visualization}. 
%The dependence of the BDW order on $U$ in the SC state is not monotonous; this is different from the normal state behavior.

The paper is organized as follows. In section \ref{sec_model}, we briefly describe the one-band Hubbard model and how the 
density waves are incorporated within this model. 
In section~\ref{sec_method}, we describe the CDMFT procedure. 
We show the important results of this work in section~\ref{sec_results} and discuss various implications 
in section \ref{sec_discussion}, and finally conclude.

%===============================================================================
\section{Model}\label{sec_model}
We use the one-band Hubbard model:

\begin{eqnarray}\label{eq:lattice_ham}
H = -\sum_{\textbf{r},\textbf{r}',\sigma}t_{\textbf{\scriptsize rr}'}\crs^{\dagger}\crs['] + 
U\sum_{\textbf{r}}n_{ \textbf{\scriptsize r},\scriptsize\uparrow}n_{ \textbf{\scriptsize r},\scriptsize\downarrow}
-\mu\sum_{\textbf{r},\sigma}n_{ \textbf{\scriptsize r},\scriptsize\sigma}
\end{eqnarray} 

where $c_{\textbf{r}{\scriptsize\sigma}}^{\dagger}$ creates an electron of spin $\sigma$ 
at the site $\textbf{r}$; $U$ is the on-site 
Coulomb repulsion and $\mu$ is the chemical potential. 
We consider only the first, second and third nearest neighbor hopping terms,
with amplitudes $t$, $t'$ and $t''$ respectively. 
We adopt the values $t'/t=-0.3$, $t''/t=0.2$, unless otherwise stated. These values of the band 
parameters are known to be appropriate for BSCO
\cite{liechtenstein1996quasiparticle} and, to some extent, for YBCO \cite{andersen1995lda}. Note that 
all energies ($U$, $\mu$, etc.) are measured in the units of $t$.

\begin{figure}[h]
\centering
\includegraphics[scale=0.1]{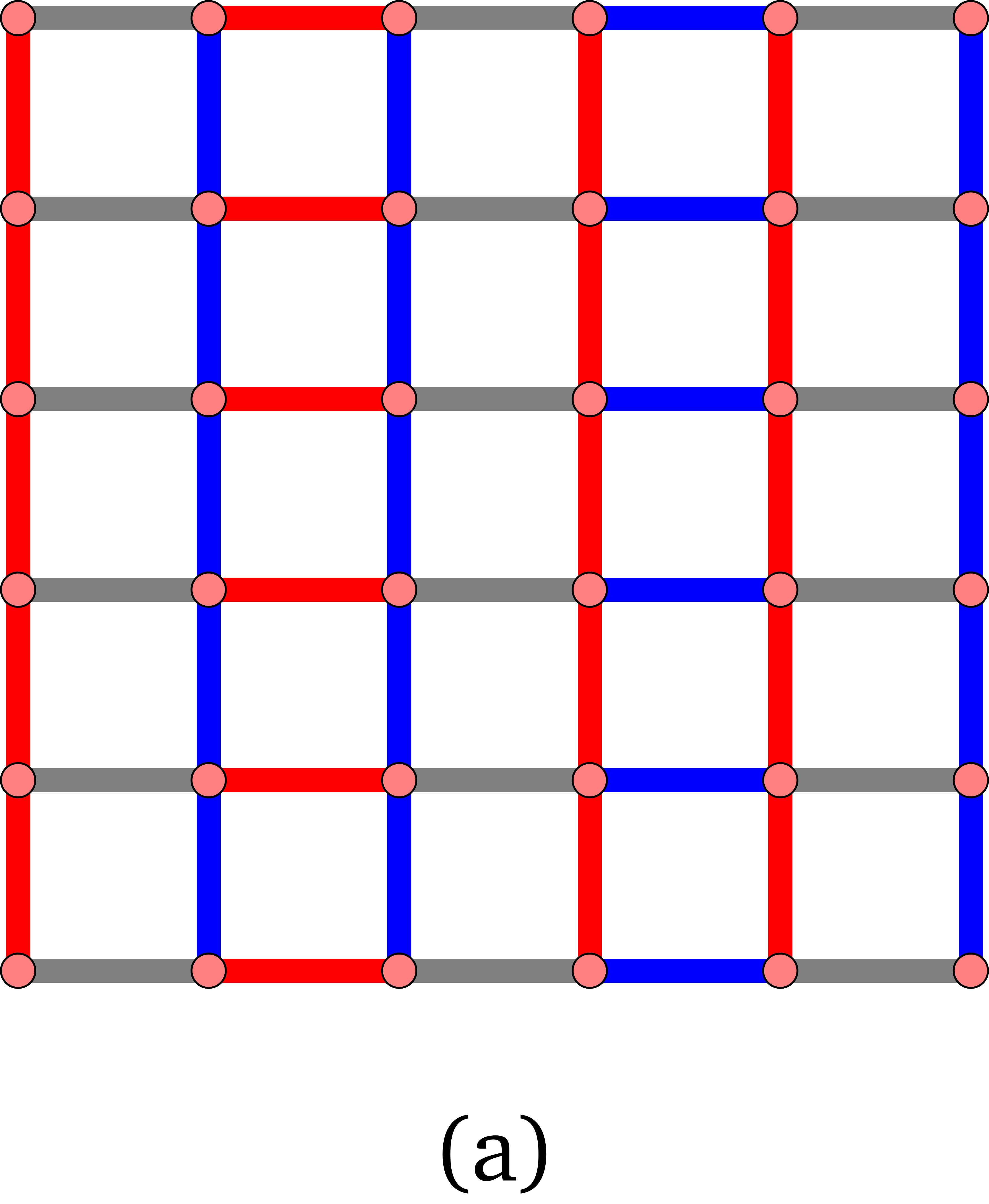}
\hspace{0.2cm}
\includegraphics[scale=0.1]{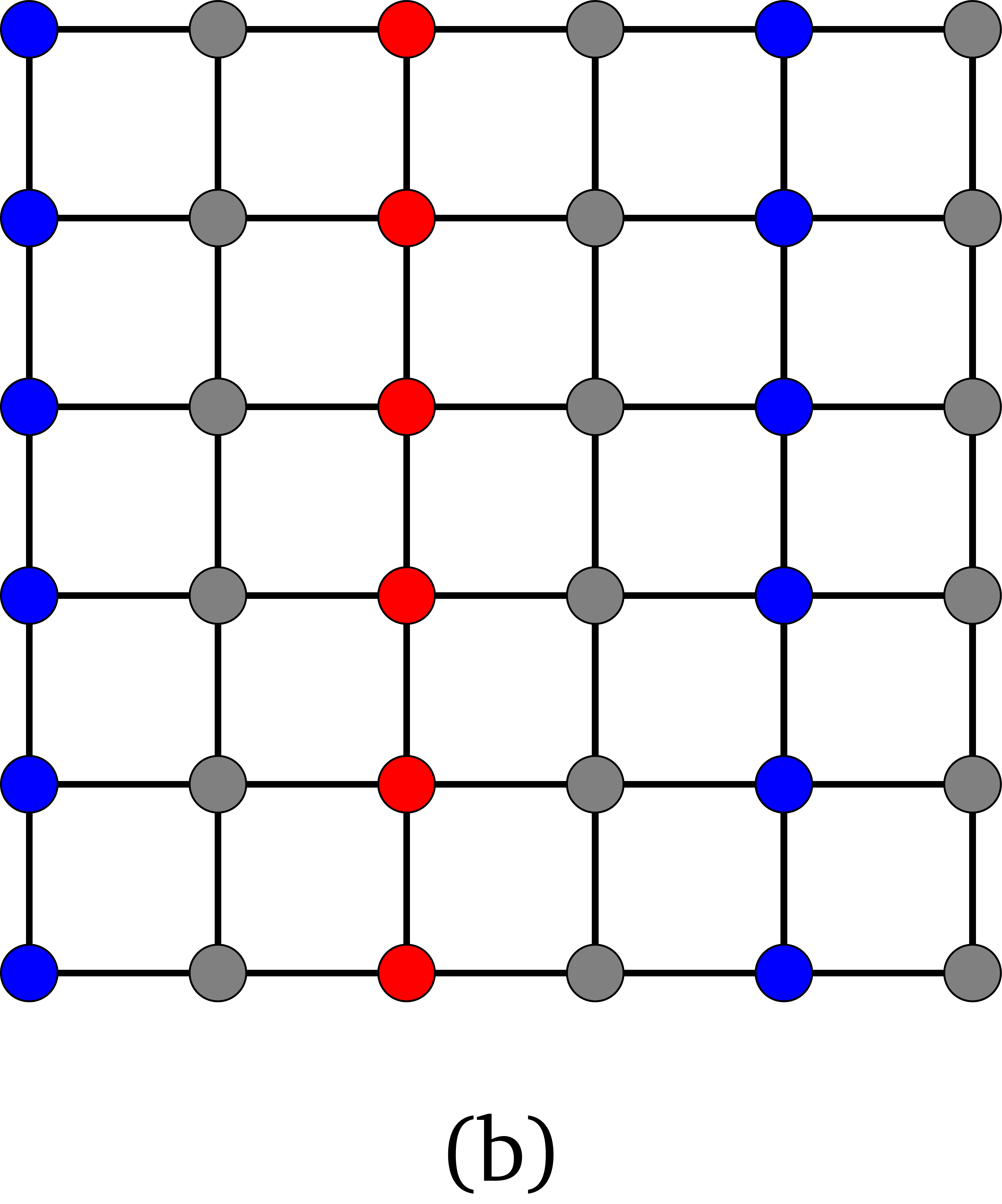}
\caption{(a) $d$-wave BDW (PDW) on the lattice (red indicates the negative amplitude, 
blue indicates the positive amplitude and gray indicates zero). 
A $s'$-wave modulation can be visualized by swapping 
the colors of the bonds (red $\rightarrow$ blue, blue $\rightarrow$ red) along any one of the axes. (b) $s$-wave CDW on the Cu sites, color code is same as in (a). } %In this work, we focus on a $d$-wave BDW modulation and a $s'$-wave PDW modulation.}
\label{fig:BDW_lattice}
\end{figure}

The charge-density modulations exist on the oxygen atoms in the CuO$_2$ lattice 
\cite{fujita2014direct,kohsaka2007intrinsic}. In the one-band 
Hubbard model, since oxygen atoms are not explicitly present, the CDWs are best represented as 
bond-density-waves (BDWs) on the Cu-Cu bonds (Fig. \ref{fig:BDW_lattice}(a)).

%Figure 1.............................................................................................................
%.............................................................................................................

The BDW operator can be written as:
\begin{equation}\label{eq:BDW_op}
\hat{\Psi}_{\textrm{BDW}}=\sum_{\rv\sigma,\av}t_{\qv,\av}c^{\dagger}_{\rv,\sigma}
c_{\rv+\av,\sigma}e^{i\qv\cdot(\rv+\av/2)} + \mathrm{H.c.}  
\end{equation}
where $t_{\qv,\xh}=-t_{\qv,\yh}$
leads to a $d$-wave form factor and 
$t_{\qv,\xh}=t_{\qv,\yh}$
leads to a $s'$-wave form factor; $\av=\pm\xh$, $\pm\yh$; since we are 
looking for modulations on the first-neighbor bonds (Fig. \ref{fig:BDW_lattice}(a)) where oxygen atoms should lie.

The $s$-wave CDW operator corresponding to the charge-density modulations 
centered on the Cu sites (Fig. \ref{fig:BDW_lattice}(b)) is written as:
\begin{equation}\label{eq:CDW_op}
\hat{\Psi}_{\textrm{CDW}}=\sum_{\rv\sigma} c^{\dagger}_{\rv,\sigma}
c_{\rv,\sigma}e^{i\qv\cdot\rv} + \mathrm{H.c.}   
\end{equation}

We also probe $d$-wave superconductivity, with the pairing operator:
\begin{multline}\label{eq:SC_op}
\hat\Psi_{\textrm{dSC}} = \sum_\rv\left(c_{\rv,\uparrow}c_{\rv\pm\xh,\downarrow}
-c_{\rv,\downarrow}c_{\rv\pm\xh,\uparrow}\right) \\
-\sum_\rv\left(c_{\rv,\uparrow}c_{\rv\pm\yh,\downarrow}
-c_{\rv,\downarrow}c_{\rv\pm\yh,\uparrow}\right) + \mathrm{H.c.}~,
\end{multline}
as well as the pair-density-wave (PDW) order with the following operator:
\begin{equation}\label{eq:PWD_op}
\hat{\Psi}_{\textrm{PDW}}=\sum_{\rv,\av}u_{\qv,\av}\left(c_{\rv,\uparrow}
c_{\rv+\av,\downarrow}-c_{\rv,\downarrow}c_{\rv+\av,\uparrow}
\right)e^{i\qv\cdot(\rv+\av/2)} + \mathrm{H.c.}  
\end{equation}   
where $u_{\qv,\xh}=u_{\qv,\yh}=1$ in a $s'$-wave form factor and $\av=\pm\xh$, $\pm\yh$. 
Motivated by experiments~\cite{kohsaka2007intrinsic,fujita2014direct}, we take $\mathbf{q}=2\pi/4\hat{\mathbf{x}}$ 
for all the above DW operators (Eqs \ref{eq:BDW_op},\ref{eq:CDW_op},\ref{eq:PWD_op}).

%===============================================================================
\section{Method}\label{sec_method}

%-------------------------------------------------------------------------------
\subsection{Cluster Dynamical Mean Field Theory}

Short-range quantum fluctuations arising from the strong local Coulomb repulsion are believed to cause the exotic orders mentioned above (Eqs \ref{eq:BDW_op}-\ref{eq:PWD_op}).
Local approaches involving one-electron excitations and formulated in terms of Green functions, such as cluster extensions of dynamical mean field theory, are known to capture these effects well.\\ 

In cluster dynamical mean field theory \cite{kotliar2001cellular}, 
the self-energy of the system is approximated by that of a self-consistent 
\textit{impurity model}, defined on a small cluster of atoms hybridized with a bath of uncorrelated orbitals.
The latter represent the effect of the cluster's environment, i.e., the rest of the lattice.
The bath parameters are adjusted self-consistently in such a way that the self-energy of the impurity problem is as close as possible to that of the infinite system.
The infinite lattice is tiled into identical, repeated units, i.e., a superlattice of identical clusters is defined, and the cluster coincides with the impurity problem.
However, in the present problem, the unit cell of the superlattice is too large to constitute a single impurity problem, and therefore two different impurity problems, each defined on a 4-site plaquette (the cluster), will be necessary to form an 8-site super unit cell (Fig.~\ref{fig:BDW_2_clus}).
We use exact diagonalization to solve each impurity model at $T=0$.
More information on this approach can be found in Refs.~\cite{kancharla2008anomalous,senechal2010bath,senechal2012cluster}.

Specifically, the impurity model (or cluster) Green function is computed:
\begin{equation}\label{eq:Dyson}
\Gv_{c}(\omega)^{-1}=\omega-\tv_c-\Gammav(\omega)-\Sigmav(\omega)
\end{equation}
where $\tv_c$ is the hopping matrix on the cluster, $\Sigmav(\omega)$ 
is the self-energy of the cluster and $\Gammav(\omega)$ is the (known)
\textit{hybridization function}, which depends on the energies of the uncorrelated orbitals and their hybridization with the cluster, aka the \textit{bath parameters}.

In the case of a single cluster per super unit cell, the lattice Green function can be written as
\begin{equation}\label{eq:lattice_G}
\Gv(\kvt , \omega)^{-1}= \omega - \tv(\kvt)-\Sigmav (\omega)
\end{equation}
where $\Sigmav(\omega)$ is the self-energy of the cluster obtained from 
\eqref{eq:Dyson} and is an approximant to the lattice self-energy, and $\kvt$ belongs to the Brillouin zone of the superlattice, aka the \textit{reduced Brillouin zone}.
$\Gv(\kvt,\omega)$ is a matrix of order $2L$, $L$ being the number of sites in the super unit cell and the factor of 2 accounting for spin.
The CDMFT self-consistency condition states that the local (projected on the cluster) Green function obtained by Fourier transforming the lattice Green function:
\begin{equation}\label{eq:Gbar}
\bar\Gv(\omega) = \frac1N \sum_{\kvt}\Gv(\kvt, \omega)
\end{equation}
should coincide with the cluster Green function $\Gv_{c}(\omega)$.
Because of the finiteness of the bath in the exact diagonalization method, this condition cannot be satisfied exactly and is instead approximated by the minimization of a distance function 
\begin{equation}\label{eq:distance}
d = \sum_ {\omega_n}W(z)\rm{Tr}\left|\mathbf{G}_{c}^{-1}(i\omega_n)-\bar\Gv^{-1}(i\omega_n)\right|^2 \;,
\end{equation}
where the sum is taken over Matsubara frequencies associated with an effective temperature, in order to avoid problems related to the discreteness of the poles in the zero-temperature Green function.
When more than one impurity model is needed in the super unit cell, some of the above formulas need to be adapted, as indicated in the next subsection.

%-------------------------------------------------------------------------------
\subsection{Choice of clusters}

The size of the super unit cell should be 
commensurate with the period of the DW we wish to probe, so that the 
super unit cell contains at least one period of the DW. 
In addition, the clusters should be large enough to allow for $d$-wave superconductivity to arise from quantum fluctuations within the cluster, and this means 4-site plaquettes.

%Figure 2.............................................................................................................
\begin{figure}[h]
\centering
\includegraphics[scale=0.15]{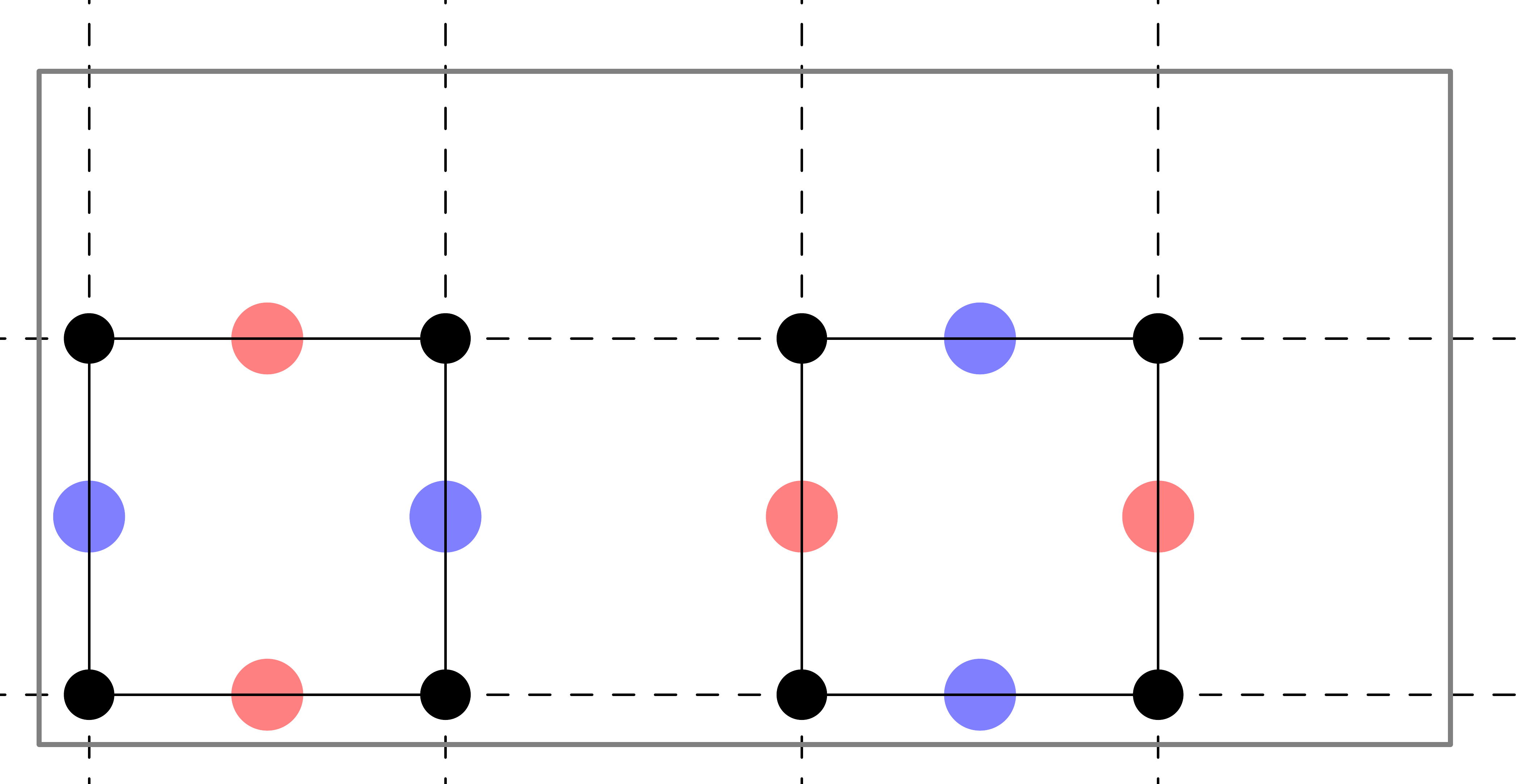}
\caption{We choose a $4a_0\times 2a_0$ super unit cell (demarcated by the box) containing two 
$2a_0\times 2a_0$ clusters. The choice is made so that a full period of 
the BDW (PDW) (shown in blue and red for positive and negative 
amplitudes respectively, on the Cu-Cu bonds) lies within the super unit cell. 
Note that we have shown a $d$-wave modulation here; a $s'$-wave modulation can be 
visualized by swapping the colors of the bonds (red $\rightarrow$ blue, blue $\rightarrow$ red) along any one of the axes.
A $s$-wave CDW (Fig. \ref{fig:BDW_lattice}(b)) is also contained within our super unit cell.}
\label{fig:BDW_2_clus}
\end{figure}
%.............................................................................................................

We therefore define a super unit cell consisting of two $2\times 2$ clusters (Fig.~\ref{fig:BDW_2_clus}). 
Each cluster forms an impurity model with a set of 8 bath orbitals, as specified below. 
With two clusters in the super unit cell, the lattice Green function takes the form 
\begin{equation}\label{eq:lattice_G_block}
\Gv(\kvt , \omega)^{-1} = \begin{pmatrix}
	\omega - \tv^{(1)}(\kvt)-\Sigmav_{1}(\omega)  &  \kern-6mm -\mathbf{t}_{\rm ic}(\mathbf{\tilde k}) \\[4pt]
-\mathbf{t}_{\rm ic}^{\dagger}(\mathbf{\tilde k})  &  \kern-6mm 	\omega - \tv^{(2)}(\kvt)-\Sigmav_{2}(\omega) 
\end{pmatrix} 
\end{equation}
where $\tv^{(i)}(\kvt)$ is the hopping matrix corresponding to cluster $i$, $\Sigmav_{i}(\omega)$ is the corresponding self-energy and
$\tv_{\rm ic}(\kvt)$ is the matrix of hoppings between clusters 1 and 2.

The distance function becomes
\begin{equation}\label{eq:dist_2_clus}
d = \sum_{j,\omega_n} W(i\omega_n)\Tr\left|\Gv_{c,j}^{-1}(i\omega_n)-\bar\Gv^{-1}_{j}(i\omega_n)\right|^2
\end{equation}
where $j$ is the cluster label. $\Gv_{c,j}$ is the $j^{\rm th}$ cluster Green function 
and $(\bar{\Gv}^{-1})_j$ is the $j^{\rm th}$ diagonal block ($2N_c\times 2N_c$) of the inverse of $\bar{\Gv}$ (Eq.~\eqref{eq:Gbar}); $N_c$ is the number of sites in each cluster. 
%-------------------------------------------------------------------------------
\subsection{Bath parametrization}\label{subsec:bath_parametrization}

The cluster-bath impurity model is defined by the Hamiltonian
\begin{eqnarray}\label{eq:imp_ham}
H_{\textrm{imp}} = H_{\textrm{clus}}+\sum_{\alpha, r,\sigma}\left(\theta_{\alpha, r} 
c_{\alpha,\sigma}^{\dagger}a_{r,\sigma} + \mathrm{H.c.}\right) 
+\sum_{r,\sigma}\epsilon_r a_{r,\sigma}^{\dagger}a_{r,\sigma}
\nonumber\\+\sum_{\beta , r}\left[\Delta_{r,\beta}\left(c_{\beta\uparrow}a_{r\downarrow}
-c_{\beta\downarrow}a_{r\uparrow}\right)+ \mathrm{H.c.}\right]\qquad\qquad\quad
\end{eqnarray}
where $H_{\textrm{clus}}$ is the restriction of Hamiltonian~\eqref{eq:lattice_ham} 
to the cluster, $c_{\alpha,\sigma}$ annihilates an electron with spin $\sigma$ at 
cluster site $\alpha$, $a_{r,\sigma}$ annihilates an electron with spin $\sigma$ in the bath orbital $r$, $\theta_{\alpha,r}$ is the hopping amplitude from bath orbital $r$ to site $\alpha$ on the cluster, $\epsilon_r$ 
is the energy of bath orbital $r$, and finally $\Delta_{r,\beta}$ is the pairing 
amplitude for a singlet formed between site $\beta$ in the cluster and bath orbital $r$. 
We have put 8 uncorrelated orbitals in the bath; this defines an impurity model of $2\times (4+8) = 24$ fermionic degrees of freedom (spin included), manageable with an ED solver.

The uncorrelated part of the impurity Hamiltonian (Eq.~\eqref{eq:imp_ham}) can 
be conveniently represented in matrix form using the Nambu formalism, i.e., in terms of the multiplet 
$(C_\uparrow, C_\downarrow^\dagger, A_\uparrow, 
A_\downarrow^\dagger)$, where $C_\s=(c_{1,\s},c_{2,\s},c_{3,\s},c_{4,\s})$ and
$A_\s=(a_{1,\s},\cdots, a_{8,\s})$ ($\s=\uparrow,\downarrow$):
\begin{equation}
H_{\mathrm{imp}}^0=\begin{pmatrix} C_{\scriptsize\uparrow}^{\dagger} &  C_{\scriptsize\downarrow} &  
A_{\scriptsize\uparrow}^{\dagger} &  A_{\scriptsize\downarrow}\end{pmatrix} \begin{pmatrix}
\mathbf{T}  &  \bm{\Theta} \\
\bm{\Theta}^{\dagger}  &  \mathbf{E}
\end{pmatrix} \begin{pmatrix} C_{\scriptsize\uparrow}\\ C_{\scriptsize\downarrow}^{\dagger}
\\ A_{\scriptsize\uparrow}\\ A_{\scriptsize\downarrow}^{\dagger}\end{pmatrix}
\end{equation}
where 
\begin{equation}
\mathbf{T} = \begin{pmatrix}
\mathbf{t}_c & \bm{0}  \\
\bm{0} & -\mathbf{t}_c\\ 
\end{pmatrix}\;,
\bm{\Theta} = \begin{pmatrix}
\thetav &  -\bm{\Delta}^{\dagger} \\
-\bm{\Delta}^{T} & -\thetav^{*}\\ 
\end{pmatrix}\;,
\mathbf{E} = \begin{pmatrix}
\epsilonv & \bm{0}  \\
\bm{0} & -\epsilonv\\ 
\end{pmatrix}
\end{equation}

$\mathbf{t}_c$ is a $4\times4$ matrix, $\thetav$ is a $4\times8$ matrix, $\bm{\Delta}$ 
is a $8\times4$ matrix, $\bm{\epsilon}$ is a $8\times 8$ diagonal matrix 
with energies of the 8 bath orbitals as the diagonal elements. \\

The uncorrelated cluster Green function can be obtained by projecting the uncorrelated impurity 
Green function $\Gv_{\mathrm{imp}}^0=\left(\bm{\omega} - H_{\mathrm{imp}}^0\right)^{-1}$ on the cluster, 
from which the bath \textit{hybridization function} $\Gammav$ in Eq.~\eqref{eq:Dyson} can be obtained as:
\begin{equation}
\Gammav = \bm{\Theta}\left(\omv - \mathbf{E}\right)^{-1}\bm{\Theta}^{\dagger}
\end{equation}

\begin{figure}[h]
\centering
\includegraphics[width=0.48\hsize]{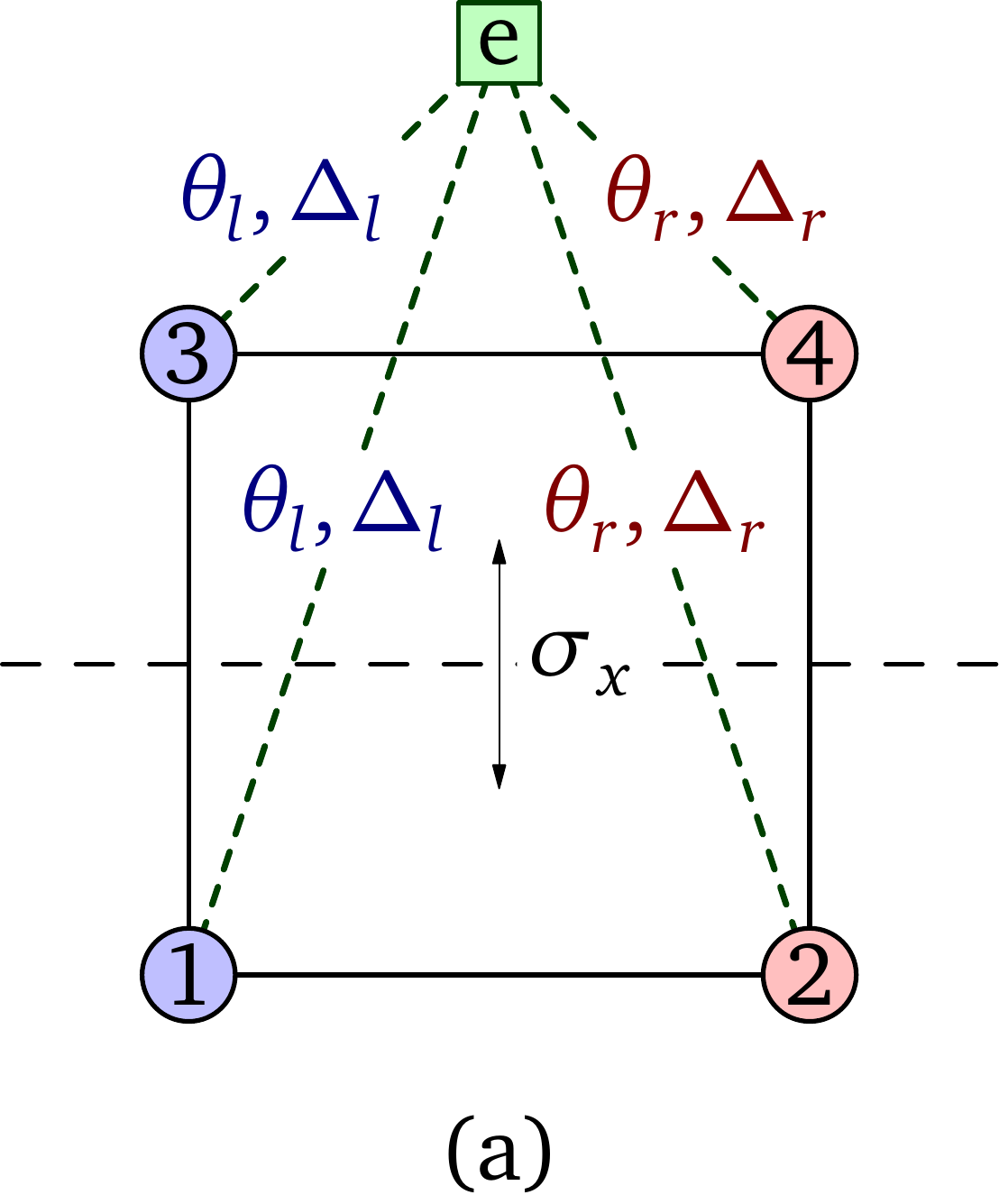}\hfill
\includegraphics[width=0.48\hsize]{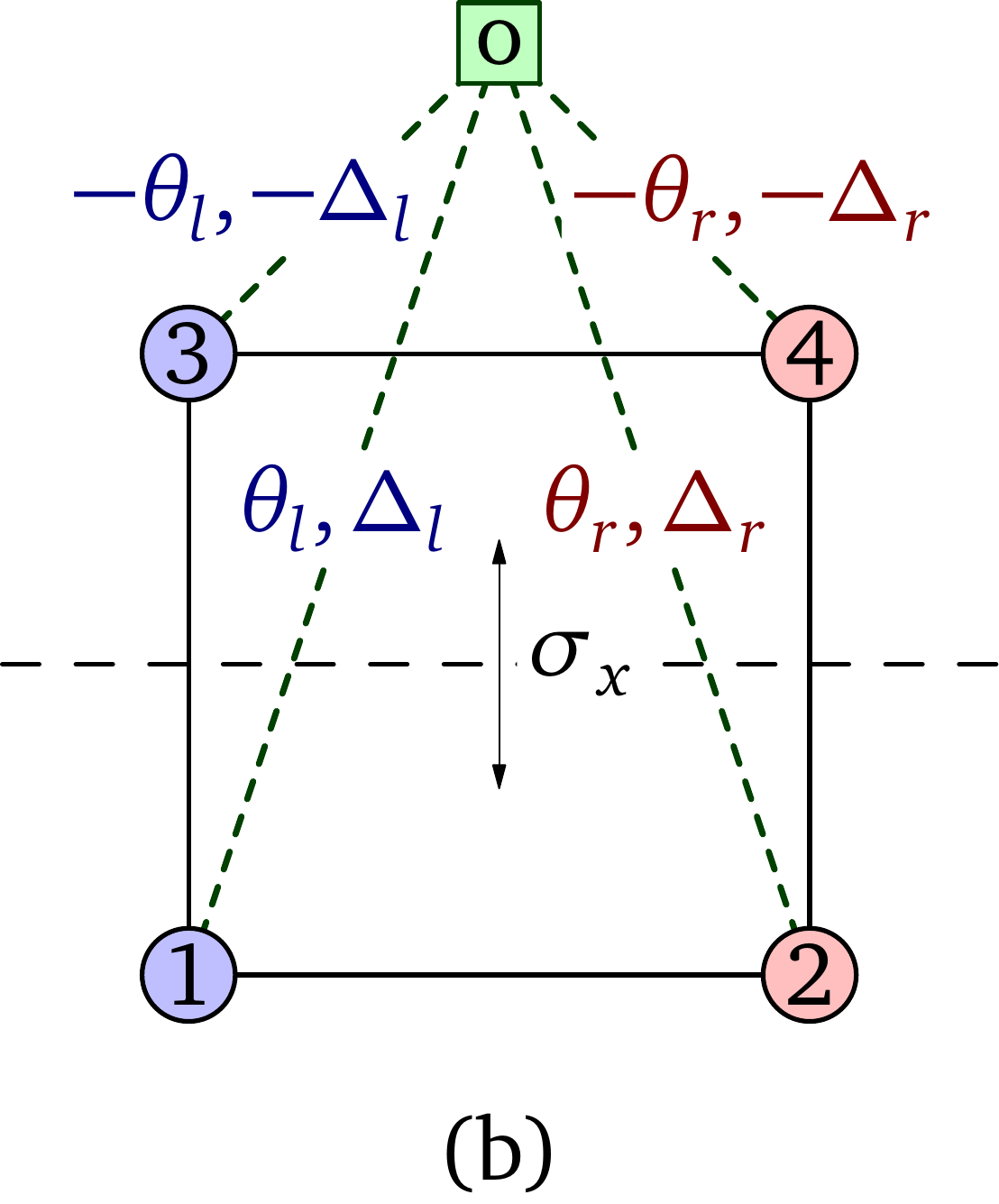}
\caption{The cluster-bath hybridization terms ($\theta' s$ and $\Delta' s$ 
in Eq.~\eqref{eq:imp_ham}) are chosen to correspond to the even and odd 
representations of the $C_2$ symmetry group, which operates on the cluster sites;  
(a) shows a cluster-bath hybridization which is even under $\sigma_{x}$ and (b) 
shows a cluster-bath hybridization which is odd under $\sigma_{x}$, 
corresponding to the two irreducible representations of $C_2$.}
\label{fig:clus_bath}
\end{figure}

There are 64 parameters in the cluster-bath hybridization $\bm{\Theta}$ for each cluster, 
therefore a total of $128$ parameters, w.r.t which the distance function 
(eq. \eqref{eq:dist_2_clus}) should be minimized at each CDMFT iteration. 
Symmetries of the cluster can help to reduce the number of independent 
variational parameters. Point group symmetries have been used to parametrize 
the bath in CDMFT \cite{koch2008sum,foley2019coexistence}. We follow Foley 
\etal~\cite{foley2019coexistence} and parametrize the bath corresponding 
to the irreducible representations of the point group $C_2$ generated by
a reflexion across the horizontal axis (Fig.~\ref{fig:clus_bath}). 
The orders that we probe, i.e., the DW and the SC orders, are compatible with this point group symmetry.

%Figure 3.............................................................................................................
%.............................................................................................................

Figure \ref{fig:clus_bath} shows the parametrization of the bath, 
which contains the even and odd irreducible representations of the 
symmetry group $C_2$. The number of independent parameters in 
$\bm{\Theta}$ is now reduced to $32$ per cluster. Out of $8$ bath orbitals, 
$4$ belong to the even irreducible representation 
(Fig.~\ref{fig:clus_bath}(a)) and 4 belong to the odd irreducible 
representation (Fig.~\ref{fig:clus_bath}(b)). The matrices $\thetav$ and 
$\bm{\Delta}$ take the following form for the even (Eqs~(\ref{eq:theta_even}-\ref{eq:Delta_even})) and odd (Eqs~(\ref{eq:theta_odd}-\ref{eq:Delta_odd})) irreducible representations:

\begin{eqnarray}
\label{eq:theta_even}\thetav_{\textrm{even}} = \begin{pmatrix}
\theta_{l1} & \theta_{l2} & \theta_{l3} & \cdots & \theta_{l8}\\
\theta_{r1} & \theta_{r2} & \theta_{r3} & \cdots & \theta_{r8}\\
\theta_{l1} & \theta_{l2} & \theta_{l3} & \cdots & \theta_{l8}\\
\theta_{r1} & \theta_{r2} & \theta_{r3} & \cdots & \theta_{r8}\\
\end{pmatrix}
\qquad\qquad\\
\label{eq:Delta_even}\bm{\Delta}_{\textrm{even}} = \begin{pmatrix}
\Delta_{l1} & \Delta_{l2} & \Delta_{l3} & \cdots & \Delta_{l8}\\
\Delta_{r1} & \Delta_{r2} & \Delta_{r3} & \cdots & \Delta_{r8}\\
\Delta_{l1} & \Delta_{l2} & \Delta_{l3} & \cdots & \Delta_{l8}\\
\Delta_{r1} & \Delta_{r2} & \Delta_{r3} & \cdots & \Delta_{r8}\\
\end{pmatrix}
\qquad\quad\\
\label{eq:theta_odd}\thetav_{\textrm{odd}} = \begin{pmatrix}
\theta_{l1} & \theta_{l2} & \theta_{l3} & \cdots & \theta_{l8}\\
\theta_{r1} & \theta_{r2} & \theta_{r3} & \cdots & \theta_{r8}\\
-\theta_{l1} & -\theta_{l2} & -\theta_{l3} & \cdots & -\theta_{l8}\\
-\theta_{r1} & -\theta_{r2} & -\theta_{r3} & \cdots & -\theta_{r8}\\
\end{pmatrix}
\quad\\
\label{eq:Delta_odd}\bm{\Delta}_{\textrm{odd}} = \begin{pmatrix}
\Delta_{l1} & \Delta_{l2} & \Delta_{l3} & \cdots & \Delta_{l8}\\
\Delta_{r1} & \Delta_{r2} & \Delta_{r3} & \cdots & \Delta_{r8}\\
-\Delta_{l1} & -\Delta_{l2} & -\Delta_{l3} & \cdots & -\Delta_{l8}\\
-\Delta_{r1} & -\Delta_{r2} & -\Delta_{r3} & \cdots & -\Delta_{r8}\\
\end{pmatrix}
\end{eqnarray}
%.............................................................................................................
Note that the bath parametrization defined by Eqs~(\ref{eq:theta_even}, \ref{eq:Delta_even}, \ref{eq:theta_odd}, \ref{eq:Delta_odd}) is 
appropriate to a solution with both DW and SC orders. 
In the normal (i.e., non-superconducting) phase, we set the cluster-bath pairing matrix $\bm{\Delta} = \bm{0}$. 
In the pure SC phase, without DW orders, the above bath parametrization is used with the constraint that $\thetav$ and $\bm{\Delta}$ are the same for both clusters.
%-------------------------------------------------------------------------------
\subsection{Computing averages}

After the CDMFT procedure has converged, the lattice Green function \eqref{eq:lattice_G_block} can be used  to compute various observables. 
The average value of a one-body operator $\hat{O}=\sum_{\rv, \rv', \sigma} O_{\rv\rv'}c^{\dagger}_{\rv\sigma}c_{\rv'\sigma}$ is computed as 
\begin{equation}
\label{lattice_average}\langle \hat{O} \rangle = 
\oint\frac{d\omega}{2\pi}\int\frac{d^2\tilde{\mathbf{k}}}{(2\pi)^2}
\rm{tr}\left[
\mathbf{O}(\tilde{\mathbf{k}})
\mathbf{G}(\tilde{\mathbf{k}},\omega)\right]
\end{equation}
called the lattice average of $\hat{O}$.
This is the formalism used to measure the order parameters corresponding 
to the operators (\ref{eq:BDW_op}-\ref{eq:PWD_op}) in this work. Another 
way of computing the averages of one-body operators is to use the cluster 
Green function (\ref{eq:Dyson}), which gives the cluster averages of operators:

\begin{equation}
	\label{cluster_average}\langle \hat{O} \rangle_c = 
	\oint\frac{d\omega}{2\pi}
	\rm{tr}\left[
	\mathbf{O}\mathbf{G}_{\textit{c}}(\omega)\right]
\end{equation}

However, the DW operators are not defined fully on each cluster because they 
are period-4 objects, hence their cluster averages would be meaningless. But for 
local operators like the density operator, it could be more relevant in some cases to 
look at the cluster averages than the lattice averages. A recent work by 
Klett~\etal~\cite{klett2020real} demonstrates the calculation of cluster 
quantities reliably on large clusters.

It is known that CDMFT intrinsically breaks the translation symmetry of the lattice, 
and this leads to a spurious DW with a period equal to the size of the cluster 
\cite{verret2019intrinsic}. We observe such a DW with a period of 2 unit cells in 
the $\xh$ direction, but this spurious effect is very different from the period-4 
DW defined in (\ref{eq:BDW_op},\ref{eq:CDW_op},\ref{eq:PWD_op}), which is accompanied 
by a difference in the electron densities between the two clusters. %that we observe as soon as the bath parameters of the two clusters start diverging from each other.

%===============================================================================
\section{Results}\label{sec_results}

In CDMFT, the bath plays a crucial role in lowering the symmetry of the ground state. 
It is designed so that the relevant symmetry is allowed to be broken. 
In this work, we have carried out CDMFT computations with the bath configuration described in Sect.~\ref{sec_method}, in three different variations to allow for three kinds of broken symmetry states: 1) A normal phase with DW orders, where translation symmetry is allowed to be broken; 2) A coexistence phase with both DW and SC orders where both translation symmetry and U(1) gauge symmetry are allowed to be broken; 
3) A pure SC phase where only the U(1) gauge symmetry is allowed to be broken. 

\subsection{Normal phase}

%Figure 4.............................................................................................................
\begin{figure}[h]
\centering
\includegraphics[width=0.9\hsize]{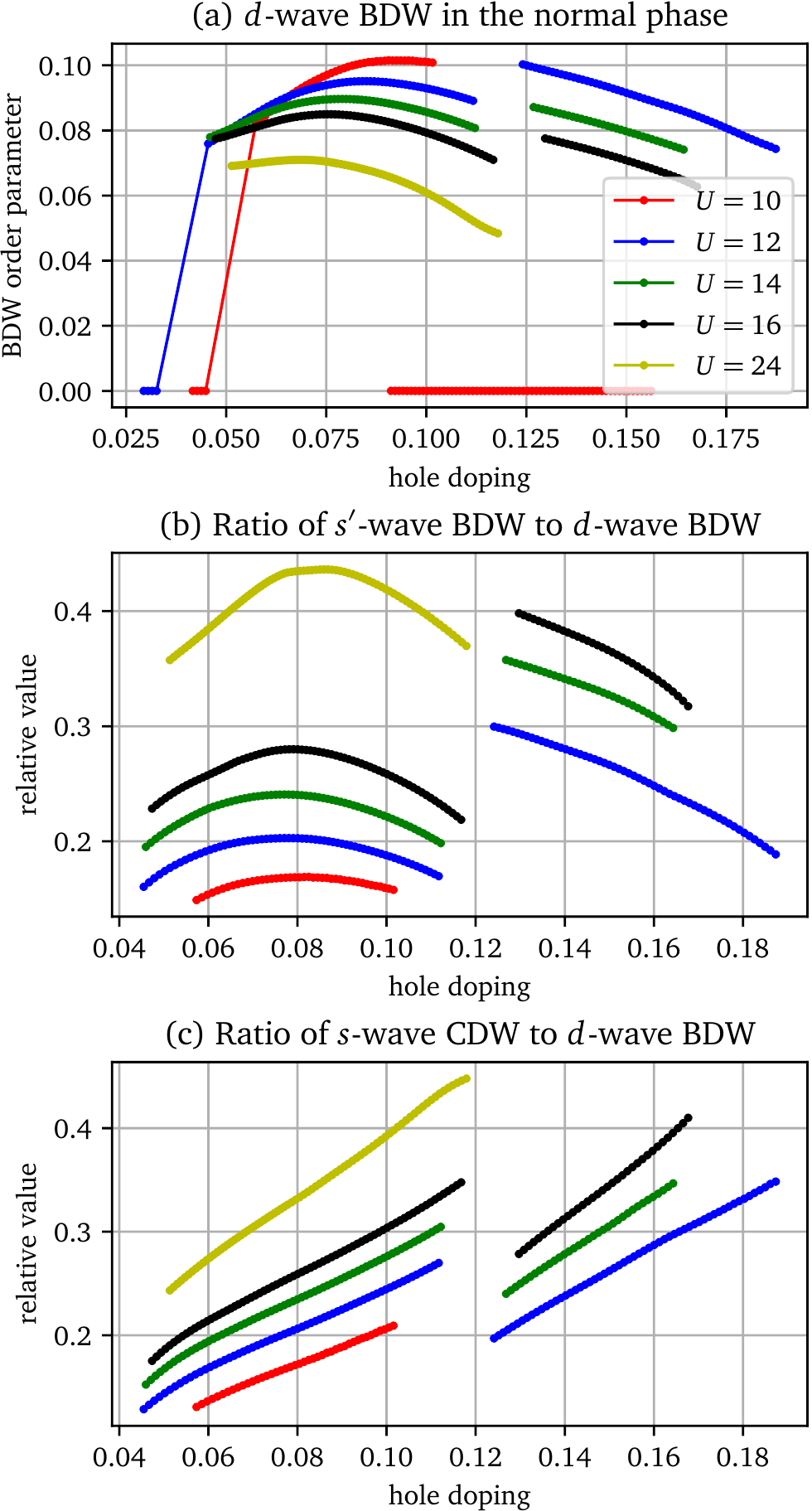}
\caption{(a) $d$-wave BDW order parameter as a function of doping for different values of $U$ in the normal phase. 
(b) $s'$-wave BDW and (c) $s$-wave CDW order parameters relative to the $d$-wave BDW order.}
\label{fig:normal_bdw}
\end{figure}
%.............................................................................................................

On doping the Mott insulator, we observe various DW orders in the normal phase, above a certain critical value of doping, for 
various values of the on-site interaction $U$. 
Figure~\ref{fig:normal_bdw} shows the $d$-wave BDW order parameter and related ($s'$- and $s$-wave) DW order parameters as a function of hole doping for different values of $U$. 
We find two disjoint solutions in the normal phase, separated by a discontinuity. 
This is an artefact of the method caused by a jump in the number of electrons in the cluster-bath system, which is a conserved quantity in normal phase computations and has occasional discontinuities upon varying the chemical potential. This is an effect of the finite size (discreteness) of the bath.

The $d$-wave BDW order parameter decreases with $U$ for most values of hole doping shown. 
The $d$-wave BDW is the predominant order, with weaker $s'$-wave BDW and $s$-wave CDW orders, as also seen in experiments 
\cite{fujita2014direct,comin2015symmetry}. 
The relative strengths of the $s'$-wave BDW and the $s$-wave CDW orders increase with $U$, 
however the $d$-wave BDW remains the dominant order for such a wide range of $U$.
Hence, we will focus on the $d$-wave BDW order for the rest of this article.

\begin{figure}[h]
\centering
\includegraphics[width=0.9\hsize]{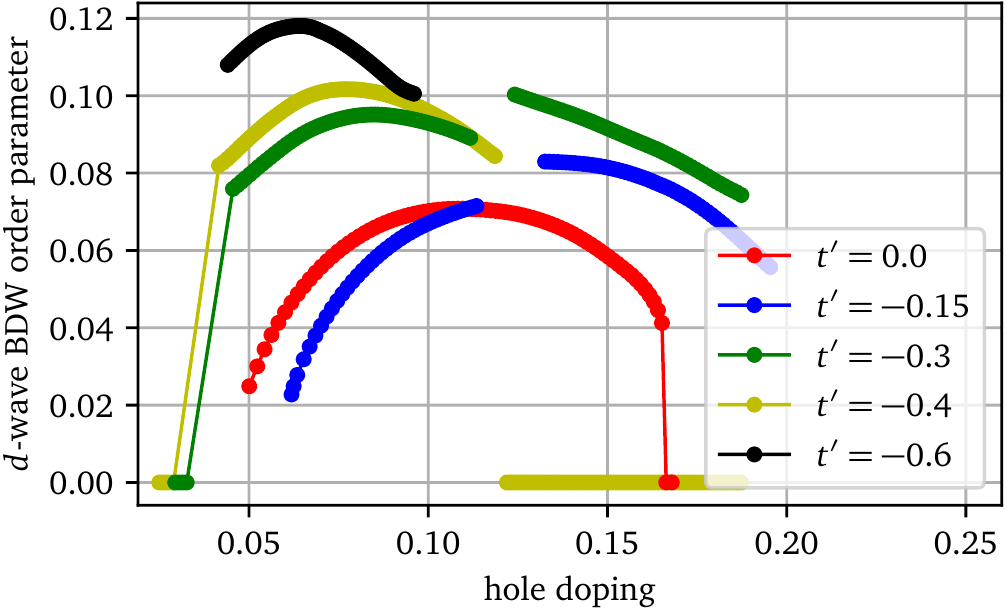}
\caption{$d$-wave BDW order parameter as a function of hole doping for different values of $t'$, at $U=12$, in 
the normal phase.}
\label{fig:normal_bdw_t1}
\end{figure}

We observe that the optimal value of the $d$-wave BDW order parameter decreases with $U$. 
This seems to support the antiferromagnetic origin of the $d$-wave BDW order.
To have a clearer indication of this, we also study the dependence of this order on the second-neighbor hopping $t'$, to see how it is affected by magnetic frustration. 
Fig.~\ref{fig:normal_bdw_t1} shows the $d$-wave BDW order parameter as a function of hole doping for different values of $t'$, at $U=12$. 
We have two disjoint solutions for most values of $t'$, like in Fig.~\ref{fig:normal_bdw}, because of a jump in the number of electrons on the impurity. 
The optimal value of the BDW order parameter increases with $|t'|$. 
This is evident for all values of $t'$ except $t'=-0.15$, in which case the order parameter is smaller than at $t'=0$ in the low doped branch; 
however it crosses the dome at $t'=0$ before reaching a maximum, which suggests that the optimal value is higher than that at 
$t'=0$. 
This suggests that the BDW order is enhanced by lattice frustration, and hence can hardly be caused by antiferromagnetic fluctuations. 

%Figure 5.............................................................................................................
%.............................................................................................................

\subsection{Coexistence phase}
\begin{figure}[h]
\centering
\includegraphics[width=0.9\hsize]{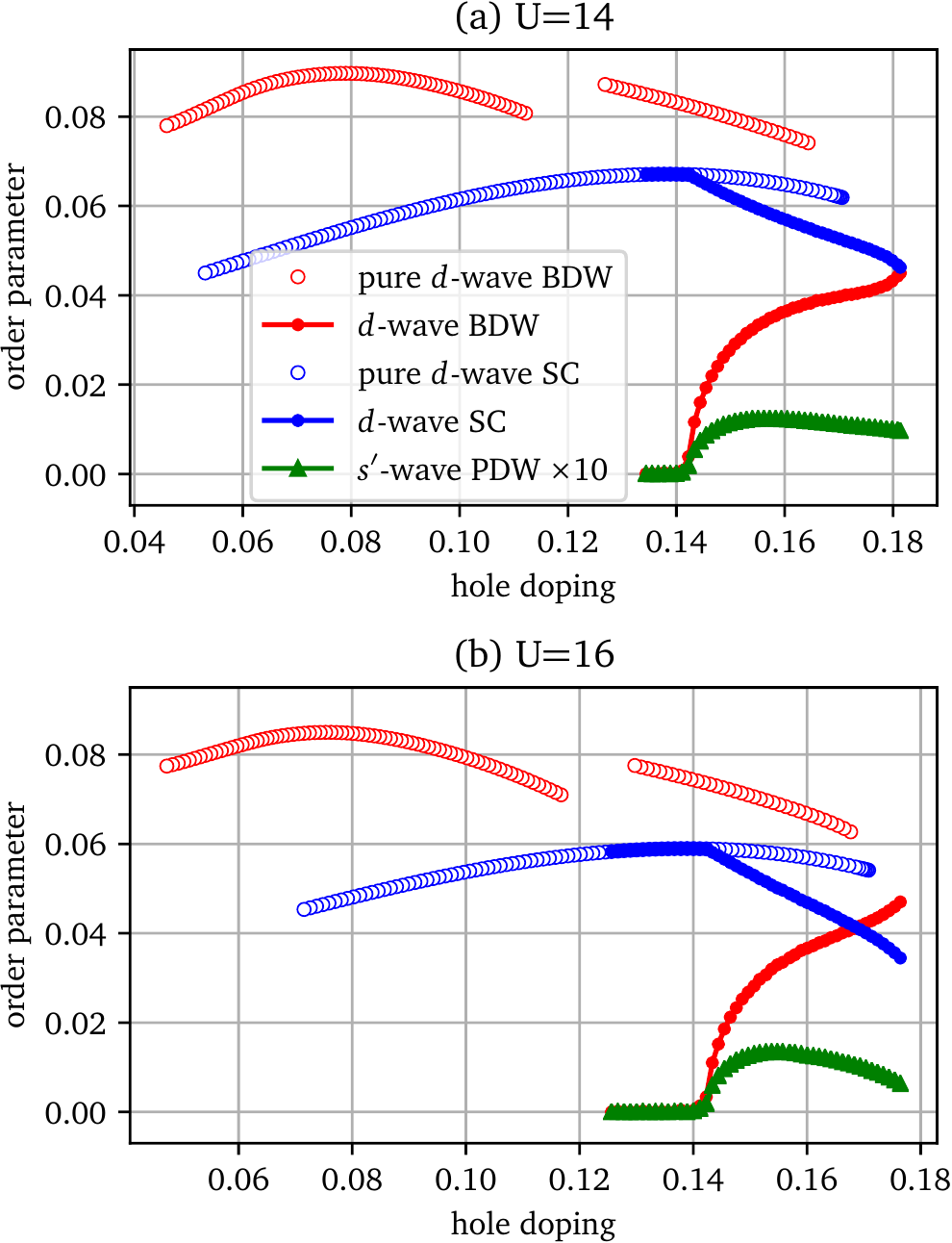}
\caption{(a) $d$-wave BDW order parameter (red) in the normal (open symbols) and coexistence (filled symbols) phases along with the SC order parameter (blue) in the pure SC (open symbols) and coexistence (filled symbols) phases, for $U=14$. The $s'$-wave PDW order (green) is also shown. (b) Same for $U=16$.}
\label{fig:bdw_dsc_pdw}
\end{figure}

To study the competition between BDW and SC orders, we use a bath parametrization that allows the two orders to exist, in what we call the coexistence phase. Here, the coexistence is microscopic, not the result of a phase separation, and hence leads to a concomitant pair-density wave (PDW). 
Fig.~\ref{fig:bdw_dsc_pdw} shows the $d$-wave BDW (filled red circles) and the SC order parameters (filled blue circles) as a function of hole doping in the coexistence phase for $U=14,\;16$. 
The BDW order starts growing at around $14\%$ doping for both values of $U$, whereas the corresponding order in the normal phase (empty red circles), exists up to much lower values of hole doping.
The magnitude of the BDW order is also much lower when it coexists with superconductivity than in the normal phase. 
This indicates that the BDW order is suppressed in the presence of superconductivity, as seen in experiments~\cite{chang2012direct,croft2014charge}. 
Moreover, the SC order parameter decreases as the BDW order parameter increases in the coexistence phase, in comparison with the pure SC phase (empty blue circles). 
Note that the value of the SC order parameter in the coexistence phase becomes equal to the value in the pure SC phase 
when the BDW order parameter decreases to zero (this is an internal consistency check). 
These signatures indicate that superconductivity is also weakened in the presence of the BDW order. 
These effects of mutual suppression indicate a competition between the two orders.

%Figure 6.............................................................................................................
%.............................................................................................................
	
\begin{figure}[h]
\centering
\includegraphics[width=0.9\hsize]{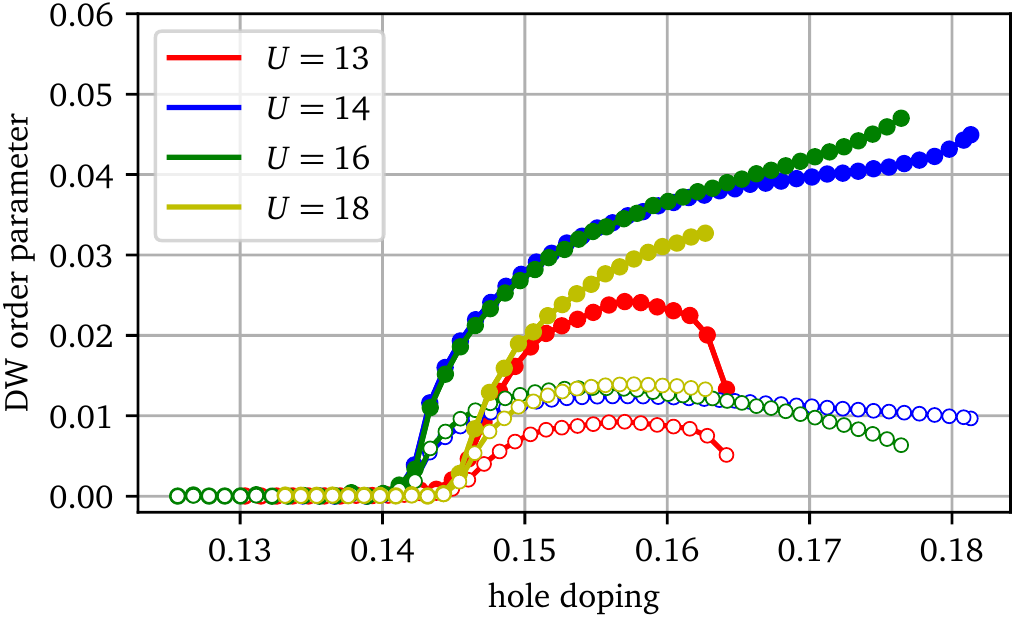}
\caption{$d$-wave BDW (filled symbols) and $s'$-wave PDW$\times 10$ (open symbols) order parameters as a function of doping, in the coexistence phase, for different values of $U$.}
\label{fig:bdw_pdw}
\end{figure}

In the coexistence region, we observe a $s'$-wave PDW order (filled green triangles in Fig. \ref{fig:bdw_dsc_pdw} 
and empty circles in Fig. \ref{fig:bdw_pdw}), as expected from the presence of both BDW and SC orders~\cite{hamidian2016detection}.
The PDW order is very weak compared to both parent orders. It is instructive to note from Fig.~\ref{fig:bdw_dsc_pdw} that the PDW order parameter decreases when either of the parent orders decrease, which confirms that it is not an independent order. 

In order to study the $U$ dependence of the DW orders in the coexistence phase, we show the BDW (filled circles) 
and the PDW (empty circles) order parameters as a function of hole doping in the coexistence phase for 
$U=13,\;14,\;16,\;18$ in Fig.~\ref{fig:bdw_pdw}. The BDW order parameter increases 
from $U=13$ to $U=14$, remains almost the same while $U$ changes from $14$ to $16$, and decreases at higher $U$ values. 
This behavior is completely different than what is observed in the normal phase (Fig.~\ref{fig:normal_bdw}), 
where the optimal value of the $d$-wave BDW order parameter decreases monotonously with $U$ for a wide range of $U$. 
One could argue that the behavior of the BDW order in the coexistence phase remains somewhat questionable for $U>13$, 
as we do not reach the optimal values for $U=14,\;16,\;18$, although this could be due to the competition with superconductivity. 
Note that the BDW order parameter tends to increase towards its normal phase value (Fig.~\ref{fig:bdw_dsc_pdw}), as the SC order parameter 
in the coexistence phase decreases. 
This might be the reason for not having a regular dome shape for the BDW order parameter here. 
However, the PDW order parameter has well-defined dome shapes and the optimal value increases from 
$U=13$ to $U=14$, and remains roughly constant thereafter.

\begin{figure}[h]
\centering
\includegraphics[width=0.9\hsize]{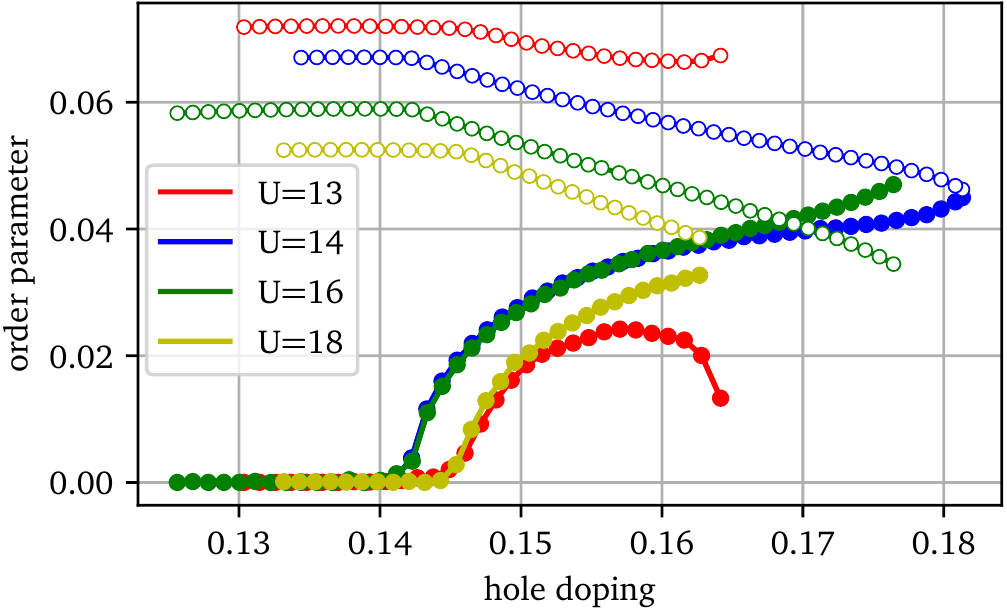}
\caption{$d$-wave BDW (filled symbols) and SC (open symbols) order parameters for $U=13,14,16,18$ in the coexistence phase.}
\label{fig:bdw_dsc}
\end{figure}

Since the BDW order is very much affected by the presence of superconductivity, it seems crucial to study the $U$ dependence of the 
BDW order in comparison with that of the SC order, as shown in Fig.~\ref{fig:bdw_dsc} (filled and empty circles respectively).
It is known that superconductivity is mediated by short-range antiferromagnetic fluctuations in the strong correlation regime within 
the one band Hubbard model, which leads $T_c^{max}$ to scale with the antiferromagnetic exchange $J$ and hence to decrease with $U$~\cite{fratino2016organizing}. 
This means that the maximum SC order parameter should decrease with $U$ when $U>U_c$~\cite{foley2019coexistence}. 
This is what we observe for the SC order parameter in the pure SC phase (empty blue circles in Figs~\ref{fig:bdw_dsc_pdw}(a), \ref{fig:bdw_dsc_pdw}(b)). 
The SC order parameter also decreases with $U$ even in the coexistence phase, keeping its behavior from the pure phase. 
However, the dependence of the BDW order on $U$ in the coexistence phase is not the same as in the normal phase, signifying that the BDW order is more affected by the coexistence than superconductivity. 
There seems to be a crossover in the behavior of the BDW order in the coexistence phase on varying $U$; it initially increases with $U$, 
and then starts decreasing at higher values of $U$. 
It is instructive to see this in the light of its competition with superconductivity. 
There are two factors which affect the BDW order in the coexistence phase: 
1) $U$: the BDW order is weakened on increasing $U$, as we know from the normal phase (Fig.~\ref{fig:normal_bdw}); 
2) superconductivity: the BDW order is weakened when the SC order grows and vice-versa (Fig.~\ref{fig:bdw_dsc_pdw}). 
When we change $U$ in the coexistence phase, the SC order also changes and the net effect on the BDW order depends on which factor dominates. 
At large values of $U$, superconductivity is weakened and BDW order follows its $U$-behavior from the normal phase, signifying that the effect of 
superconductivity does not dominate in this regime.
By contrast, on increasing $U$ at smaller values of $U$, where superconductivity is strong, the BDW order is strengthened when superconductivity is weakened with $U$, deviating from its behavior in the normal phase.  
	
%Figure 7.............................................................................................................
%.............................................................................................................

%Figure 8.............................................................................................................
%.............................................................................................................

%===============================================================================
\section{Discussion and Conclusion}\label{sec_discussion}	

We were able to obtain various period-four DW orders arising just from local correlation effects in the one-band Hubbard 
model, with the $d$-wave BDW as the dominant order in the normal phase. 
The $d$-wave BDW order coexists with superconductivity in a certain region of doping, where it is weaker than in the normal phase. 
Superconductivity is also weakened when it coexists with BDW order, indicating a competition between the two orders. 
Additionally, a $s'$-wave PDW order is observed in the region where both $d$-wave BDW and SC orders exist. 
The $d$-wave BDW order is suppressed as $U$ increases in the normal phase. 
However, the $U$-dependence of the BDW order parameter in the coexistence phase is non-monotonous; it increases first and then 
decreases above a certain $U$, indicating a crossover. 
We believe that this happens because of the competition between BDW order and superconductivity, which is significant at lower values of $U$ (where the BDW order parameter increases with $U$) but does not play a significant role at higher values of $U$ where we recover the normal phase behavior. 
Furthermore, we observe that the BDW order grows on increasing $|t'|$, which suggests that the suppression of antiferromagnetic fluctuations favors the BDW order. 

\begin{figure}[h]
\centering
\includegraphics[width=0.9\hsize]{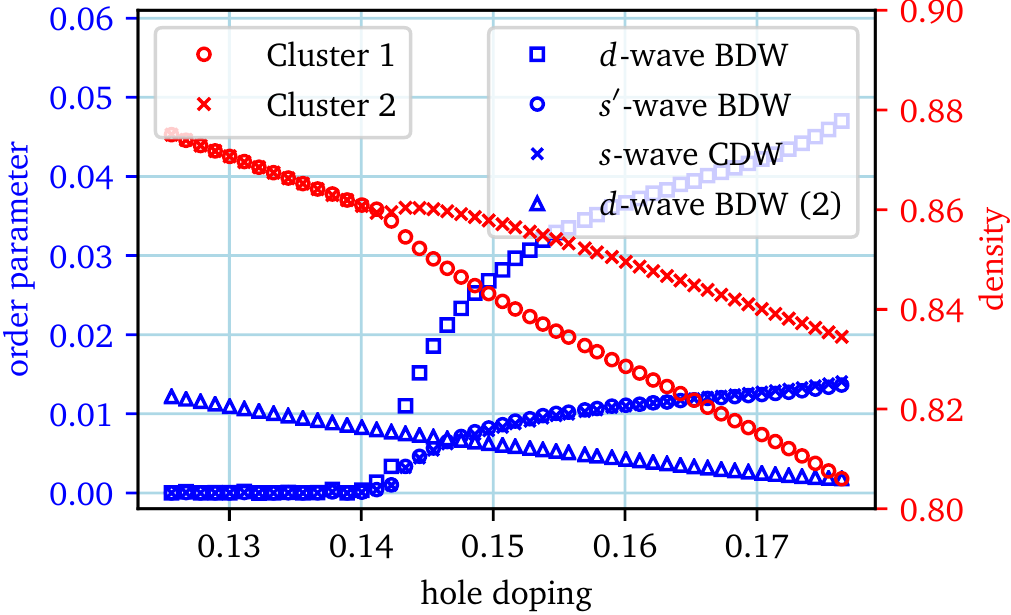}
\caption{Cluster densities (red) are shown along with the period-4 DW order parameters (blue circles, squares and crosses) 
for the coexistence phase for $U=16$. The two clusters are identical in the absence of a period-4 DW order, 
which then start to diverge on their densities as the period-4 DW orders develop. However, we always have a period-2 BDW order 
(blue triangles), which appears as an artifact of the method.}
\label{fig:nclus}
\end{figure}

The period-4 DW orders appear spontaneously in our CDMFT solutions as we reach appropriate doping values.
This is marked by a spontaneous breaking of the translation symmetry in our impurity models, i.e., across the two impurity clusters, which constitute the unit cell of the superlattice.   
For instance, the appearance of the period-4 DW orders is accompanied by a difference in electronic density between the two clusters. 
Figure~\ref{fig:nclus} shows the electron density on the two clusters (red symbols) along with the various DW order parameters (blue symbols) as a function of lattice doping. 
This difference in density between the two clusters can be attributed to the $s'$-wave BDW and the $s$-wave CDW orders, which appear along with the dominant $d$-wave BDW order, as also seen in Fig.~\ref{fig:normal_bdw}, 
since the $d$-wave BDW order should not produce a bias in the electron density in one cluster against the other. 
We also observe a period-2 BDW (blue triangles in Fig.~\ref{fig:nclus}). 
This is an artifact of the method: the smallest unit to be solved exactly is a $2\times 2$ cluster and hence the bonds within and between the cluster are treated differently. 
Such a period-2 BDW order is present almost in the entire region of hole doping explored and is not affected by the transitions in the period-4 DW orders which indicate physical phase transitions. 
% Note further that the period-4 modulations defined here (Fig.~\ref{fig:BDW_lattice}(a)) cannot be caused by two subsequent period-2 modulations, and hence are not related. 

%Figure 9.............................................................................................................
%.............................................................................................................

%The $U$-dependence of the optimal value of the $d$-wave BDW order in the normal phase (Fig.~\ref{fig:normal_bdw}) hinted at a possible
%antiferromagnetic origin of the BDW order, as the exchange constant $J$ scales as $1/U$ for large values of $U$~\cite{delannoy2005neel}. 
%more on the 1/U figure.

\begin{figure}[h]
\centering
\includegraphics[width=0.85\hsize]{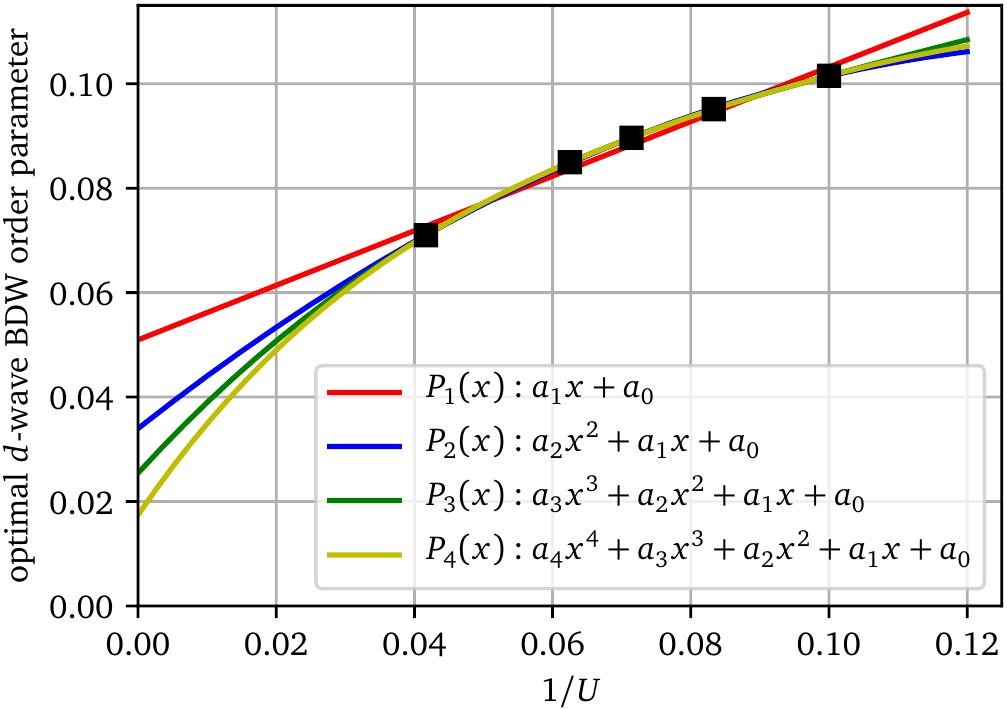}
\caption{Optimal value of the $d$-wave BDW order parameter in the 
normal phase (Fig.~\ref{fig:normal_bdw}(a)) as a function of $1/U$ along with fits 
with various polynomial functions. The values of the coefficients 
for each polynomial function is shown in table \ref{tab:fitting_parameters}.}
\label{fig:fit_J}
\end{figure}

The $U$-dependence of the $d$-wave BDW order in the normal phase (Fig.~\ref{fig:normal_bdw}) is shown in Fig.~\ref{fig:fit_J}, 
where we have plotted the optimal value of the order parameter at each $U$, as a function of $1/U$. The plot is 
also accompanied by fits of the data with polynomial functions in $1/U$ of degrees $1$ to $4$. The key observation is that 
the constant term $a_0$ (see table~\ref{tab:fitting_parameters}) decreases as the degree of the polynomial increases, which indicates that the optimal order parameter goes 
to zero as $U\rightarrow \infty$. This hints at a possible antiferromagnetic origin of the BDW order, since the exchange 
correlation $J$ scales as $1/U$ for large values of $U$.
However, the dependence on $t'$ (Fig.~\ref{fig:normal_bdw_t1}) suggests otherwise. 
The optimal value of the $d$-wave BDW order parameter increases with $|t'|$, indicating that lattice frustration favors the BDW order. 
To strengthen this interpretation, one needs to make sure that the increase in the BDW order parameter with $|t'|$ is not due to a decrease in 
the effective interaction (ratio of $U$ over bandwidth of the non-interacting density of states). 
To this end, we checked that the bandwidth of the non-interacting density of states remains at a constant value of $8$ as $t'$ changes from $0$ to $-0.4$. 
However, the bandwidth increases to $8.8$ as $t'$ changes to $-0.6$. 
Hence our conclusion that the BDW order is favored by lattice frustration remains valid, since the BDW order parameter increases as $t'$ changes from $0$ to $-0.4$ with a constant non-interacting bandwidth. 
The growth of the BDW order as $t'$ changes from $-0.4$ to $-0.6$ could be due to the combined effect of a larger lattice frustration 
and a smaller effective interaction. We know that lattice frustration suppresses antiferromagnetic fluctuations, and the fact that it 
favors the BDW order could mean that antiferromagnetism and BDW order suppress each other. 
This could be a sign of a competition between BDW order and antiferromagnetism~\cite{bauer2010competition}.

%Figure 10.............................................................................................................

It is interesting to compare our key observations with those of a recent study of 
the interplay between BDW and SC orders by Faye and S\'en\'echal~\cite{faye2017interplay}. 
In their work, they observe a BDW order alone and in presence of superconductivity, 
arising out of local correlation effects within the variational cluster approximation 
(VCA) in the one-band Hubbard model. 
They observe a growth of the BDW order with $U$ around the Mott transition point. 
In this regime, the on-site coulomb interaction seems to favor the BDW order. 
On the other hand, we study the BDW order in the strong coupling regime, 
where it is weakened by $U$. 
They also observe that the BDW order and superconductivity suppress each 
other when they coexist and lead to a weaker PDW order.
This is consistent with what we observe. Further, the overall $U$-dependence of the BDW 
order is in qualitative agreement with a large-$N$ mean-field study of the $t$-$J$ 
model~\cite{vojta2000competing,vojta2002superconducting}. Vojta~\cite{vojta2002superconducting} 
observed the onset of a charge-ordered state below a critical value of $J$ and further found that 
charge ordering is weakened as $J$ decreases, at small values of $J$.

To summarize, we observe a dominant $d$-wave BDW order in the one-band Hubbard model, both in the normal phase and in coexistence with superconductivity, 
along with weaker $s$ and $s'$-wave DW orders, using CDMFT with an ED impurity solver. 
The $d$-wave BDW order competes with superconductivity in the coexistence region, where it also leads to a $s'$-wave PDW order. 
The $d$-wave BDW order is suppressed as $U$ increases in the strong correlation regime. 
Finally, the BDW order grows with $|t'|$, indicating that it is favored by lattice frustration and competes with antiferromagnetism.   

\begin{table}[]
	\caption{Fit parameters of the polynomial functions used in Fig.~\ref{fig:fit_J}.}\label{tab:fitting_parameters}
	\begin{tabular}{l|l|l|l|l}
	%\hline
	&  $P_1(x)$  &  $P_2(x)$  &  $P_3(x)$  &  $P_4(x)$ \\ \toprule
	$a_0$  &     $0.051\pm0.0031$   &  $0.034\pm0.0012$    &  $0.025\pm0.00077$   &   $0.017$ \\ \hline
	$a_1$  &    $0.52\pm0.042$   &  $1.043\pm0.034$    &   $1.45\pm0.036$  &   $1.94$   \\ \hline
	$a_2$  &        & $-3.68\pm0.24$  &   $-9.71\pm0.54$  &  $-20.73$   \\ \hline
	$a_3$  &        &      &   $28.35\pm2.52$  &  $133.67$  \\ \hline
	$a_4$  &        &      &     &   $-366.26$ \\    
	\end{tabular}
\end{table}

\acknowledgments
Fruitful discussions with A. Foley, S. Verret and A.-M.~Tremblay are gratefully acknowledged.
Computational resources for this work were provided by Compute Canada and Calcul Qu\'ebec. 
This work has been supported by the Natural Sciences and Engineering Research Council of Canada (NSERC) 
under grants RGPIN-2015-05598 and RGPIN-2020-05060.

%==========================================================================================
%\bibliography{references.bib}

\begin{thebibliography}{49}%
	\makeatletter
	\providecommand \@ifxundefined [1]{%
	 \@ifx{#1\undefined}
	}%
	\providecommand \@ifnum [1]{%
	 \ifnum #1\expandafter \@firstoftwo
	 \else \expandafter \@secondoftwo
	 \fi
	}%
	\providecommand \@ifx [1]{%
	 \ifx #1\expandafter \@firstoftwo
	 \else \expandafter \@secondoftwo
	 \fi
	}%
	\providecommand \natexlab [1]{#1}%
	\providecommand \enquote  [1]{``#1''}%
	\providecommand \bibnamefont  [1]{#1}%
	\providecommand \bibfnamefont [1]{#1}%
	\providecommand \citenamefont [1]{#1}%
	\providecommand \href@noop [0]{\@secondoftwo}%
	\providecommand \href [0]{\begingroup \@sanitize@url \@href}%
	\providecommand \@href[1]{\@@startlink{#1}\@@href}%
	\providecommand \@@href[1]{\endgroup#1\@@endlink}%
	\providecommand \@sanitize@url [0]{\catcode `\\12\catcode `\$12\catcode
	  `\&12\catcode `\#12\catcode `\^12\catcode `\_12\catcode `\%12\relax}%
	\providecommand \@@startlink[1]{}%
	\providecommand \@@endlink[0]{}%
	\providecommand \url  [0]{\begingroup\@sanitize@url \@url }%
	\providecommand \@url [1]{\endgroup\@href {#1}{\urlprefix }}%
	\providecommand \urlprefix  [0]{URL }%
	\providecommand \Eprint [0]{\href }%
	\providecommand \doibase [0]{http://dx.doi.org/}%
	\providecommand \selectlanguage [0]{\@gobble}%
	\providecommand \bibinfo  [0]{\@secondoftwo}%
	\providecommand \bibfield  [0]{\@secondoftwo}%
	\providecommand \translation [1]{[#1]}%
	\providecommand \BibitemOpen [0]{}%
	\providecommand \bibitemStop [0]{}%
	\providecommand \bibitemNoStop [0]{.\EOS\space}%
	\providecommand \EOS [0]{\spacefactor3000\relax}%
	\providecommand \BibitemShut  [1]{\csname bibitem#1\endcsname}%
	\let\auto@bib@innerbib\@empty
	%</preamble>
	\bibitem [{\citenamefont {Ghiringhelli}\ \emph {et~al.}(2012)\citenamefont
	  {Ghiringhelli}, \citenamefont {Le~Tacon}, \citenamefont {Minola},
	  \citenamefont {Blanco-Canosa}, \citenamefont {Mazzoli}, \citenamefont
	  {Brookes}, \citenamefont {De~Luca}, \citenamefont {Frano}, \citenamefont
	  {Hawthorn}, \citenamefont {He} \emph {et~al.}}]{ghiringhelli2012long}%
	  \BibitemOpen
	  \bibfield  {author} {\bibinfo {author} {\bibfnamefont {G}~\bibnamefont
	  {Ghiringhelli}}, \bibinfo {author} {\bibfnamefont {M}~\bibnamefont
	  {Le~Tacon}}, \bibinfo {author} {\bibfnamefont {Matteo}\ \bibnamefont
	  {Minola}}, \bibinfo {author} {\bibfnamefont {S}~\bibnamefont
	  {Blanco-Canosa}}, \bibinfo {author} {\bibfnamefont {Claudio}\ \bibnamefont
	  {Mazzoli}}, \bibinfo {author} {\bibfnamefont {NB}~\bibnamefont {Brookes}},
	  \bibinfo {author} {\bibfnamefont {GM}~\bibnamefont {De~Luca}}, \bibinfo
	  {author} {\bibfnamefont {A}~\bibnamefont {Frano}}, \bibinfo {author}
	  {\bibfnamefont {DG}~\bibnamefont {Hawthorn}}, \bibinfo {author}
	  {\bibfnamefont {F}~\bibnamefont {He}},  \emph {et~al.},\ }\bibfield  {title}
	  {\enquote {\bibinfo {title} {Long-range incommensurate charge fluctuations in
	  {(Y,Nd)Ba$_2$Cu$_3$O$_{6+x}$}},}\ }\href@noop {} {\bibfield  {journal}
	  {\bibinfo  {journal} {Science}\ }\textbf {\bibinfo {volume} {337}},\ \bibinfo
	  {pages} {821--825} (\bibinfo {year} {2012})}\BibitemShut {NoStop}%
	\bibitem [{\citenamefont {Kohsaka}\ \emph {et~al.}(2007)\citenamefont
	  {Kohsaka}, \citenamefont {Taylor}, \citenamefont {Fujita}, \citenamefont
	  {Schmidt}, \citenamefont {Lupien}, \citenamefont {Hanaguri}, \citenamefont
	  {Azuma}, \citenamefont {Takano}, \citenamefont {Eisaki}, \citenamefont
	  {Takagi} \emph {et~al.}}]{kohsaka2007intrinsic}%
	  \BibitemOpen
	  \bibfield  {author} {\bibinfo {author} {\bibfnamefont {Y}~\bibnamefont
	  {Kohsaka}}, \bibinfo {author} {\bibfnamefont {C}~\bibnamefont {Taylor}},
	  \bibinfo {author} {\bibfnamefont {K}~\bibnamefont {Fujita}}, \bibinfo
	  {author} {\bibfnamefont {A}~\bibnamefont {Schmidt}}, \bibinfo {author}
	  {\bibfnamefont {C}~\bibnamefont {Lupien}}, \bibinfo {author} {\bibfnamefont
	  {T}~\bibnamefont {Hanaguri}}, \bibinfo {author} {\bibfnamefont
	  {M}~\bibnamefont {Azuma}}, \bibinfo {author} {\bibfnamefont {M}~\bibnamefont
	  {Takano}}, \bibinfo {author} {\bibfnamefont {H}~\bibnamefont {Eisaki}},
	  \bibinfo {author} {\bibfnamefont {H}~\bibnamefont {Takagi}},  \emph
	  {et~al.},\ }\bibfield  {title} {\enquote {\bibinfo {title} {An intrinsic
	  bond-centered electronic glass with unidirectional domains in underdoped
	  cuprates},}\ }\href@noop {} {\bibfield  {journal} {\bibinfo  {journal}
	  {Science}\ }\textbf {\bibinfo {volume} {315}},\ \bibinfo {pages} {1380--1385}
	  (\bibinfo {year} {2007})}\BibitemShut {NoStop}%
	\bibitem [{\citenamefont {Fujita}\ \emph {et~al.}(2014)\citenamefont {Fujita},
	  \citenamefont {Hamidian}, \citenamefont {Edkins}, \citenamefont {Kim},
	  \citenamefont {Kohsaka}, \citenamefont {Azuma}, \citenamefont {Takano},
	  \citenamefont {Takagi}, \citenamefont {Eisaki}, \citenamefont {Uchida} \emph
	  {et~al.}}]{fujita2014direct}%
	  \BibitemOpen
	  \bibfield  {author} {\bibinfo {author} {\bibfnamefont {Kazuhiro}\
	  \bibnamefont {Fujita}}, \bibinfo {author} {\bibfnamefont {Mohammad~H}\
	  \bibnamefont {Hamidian}}, \bibinfo {author} {\bibfnamefont {Stephen~D}\
	  \bibnamefont {Edkins}}, \bibinfo {author} {\bibfnamefont {Chung~Koo}\
	  \bibnamefont {Kim}}, \bibinfo {author} {\bibfnamefont {Yuhki}\ \bibnamefont
	  {Kohsaka}}, \bibinfo {author} {\bibfnamefont {Masaki}\ \bibnamefont {Azuma}},
	  \bibinfo {author} {\bibfnamefont {Mikio}\ \bibnamefont {Takano}}, \bibinfo
	  {author} {\bibfnamefont {Hidenori}\ \bibnamefont {Takagi}}, \bibinfo {author}
	  {\bibfnamefont {Hiroshi}\ \bibnamefont {Eisaki}}, \bibinfo {author}
	  {\bibfnamefont {Shin-ichi}\ \bibnamefont {Uchida}},  \emph {et~al.},\
	  }\bibfield  {title} {\enquote {\bibinfo {title} {Direct phase-sensitive
	  identification of a $d$-form factor density wave in underdoped cuprates},}\
	  }\href@noop {} {\bibfield  {journal} {\bibinfo  {journal} {Proceedings of the
	  National Academy of Sciences}\ }\textbf {\bibinfo {volume} {111}},\ \bibinfo
	  {pages} {E3026--E3032} (\bibinfo {year} {2014})}\BibitemShut {NoStop}%
	\bibitem [{\citenamefont {Comin}\ and\ \citenamefont
	  {Damascelli}(2016)}]{comin2016resonant}%
	  \BibitemOpen
	  \bibfield  {author} {\bibinfo {author} {\bibfnamefont {Riccardo}\
	  \bibnamefont {Comin}}\ and\ \bibinfo {author} {\bibfnamefont {Andrea}\
	  \bibnamefont {Damascelli}},\ }\bibfield  {title} {\enquote {\bibinfo {title}
	  {Resonant x-ray scattering studies of charge order in cuprates},}\
	  }\href@noop {} {\bibfield  {journal} {\bibinfo  {journal} {Annual Review of
	  Condensed Matter Physics}\ }\textbf {\bibinfo {volume} {7}},\ \bibinfo
	  {pages} {369--405} (\bibinfo {year} {2016})}\BibitemShut {NoStop}%
	\bibitem [{\citenamefont {Huecker}\ \emph {et~al.}(2014)\citenamefont
	  {Huecker}, \citenamefont {Christensen}, \citenamefont {Holmes}, \citenamefont
	  {Blackburn}, \citenamefont {Forgan}, \citenamefont {Liang}, \citenamefont
	  {Bonn}, \citenamefont {Hardy}, \citenamefont {Gutowski}, \citenamefont
	  {Zimmermann} \emph {et~al.}}]{huecker2014competing}%
	  \BibitemOpen
	  \bibfield  {author} {\bibinfo {author} {\bibfnamefont {Markus}\ \bibnamefont
	  {Huecker}}, \bibinfo {author} {\bibfnamefont {Niels~Bech}\ \bibnamefont
	  {Christensen}}, \bibinfo {author} {\bibfnamefont {AT}~\bibnamefont {Holmes}},
	  \bibinfo {author} {\bibfnamefont {Elizabeth}\ \bibnamefont {Blackburn}},
	  \bibinfo {author} {\bibfnamefont {Edward~M}\ \bibnamefont {Forgan}}, \bibinfo
	  {author} {\bibfnamefont {Ruixing}\ \bibnamefont {Liang}}, \bibinfo {author}
	  {\bibfnamefont {DA}~\bibnamefont {Bonn}}, \bibinfo {author} {\bibfnamefont
	  {WN}~\bibnamefont {Hardy}}, \bibinfo {author} {\bibfnamefont {Olof}\
	  \bibnamefont {Gutowski}}, \bibinfo {author} {\bibfnamefont {M~v}\
	  \bibnamefont {Zimmermann}},  \emph {et~al.},\ }\bibfield  {title} {\enquote
	  {\bibinfo {title} {Competing charge, spin, and superconducting orders in
	  underdoped {YBa$_2$Cu$_3$O$_y$}},}\ }\href@noop {} {\bibfield  {journal}
	  {\bibinfo  {journal} {Physical Review B}\ }\textbf {\bibinfo {volume} {90}},\
	  \bibinfo {pages} {054514} (\bibinfo {year} {2014})}\BibitemShut {NoStop}%
	\bibitem [{\citenamefont {Chang}\ \emph {et~al.}(2012)\citenamefont {Chang},
	  \citenamefont {Blackburn}, \citenamefont {Holmes}, \citenamefont
	  {Christensen}, \citenamefont {Larsen}, \citenamefont {Mesot}, \citenamefont
	  {Liang}, \citenamefont {Bonn}, \citenamefont {Hardy}, \citenamefont
	  {Watenphul} \emph {et~al.}}]{chang2012direct}%
	  \BibitemOpen
	  \bibfield  {author} {\bibinfo {author} {\bibfnamefont {J}~\bibnamefont
	  {Chang}}, \bibinfo {author} {\bibfnamefont {E}~\bibnamefont {Blackburn}},
	  \bibinfo {author} {\bibfnamefont {AT}~\bibnamefont {Holmes}}, \bibinfo
	  {author} {\bibfnamefont {Niels~B}\ \bibnamefont {Christensen}}, \bibinfo
	  {author} {\bibfnamefont {Jacob}\ \bibnamefont {Larsen}}, \bibinfo {author}
	  {\bibfnamefont {J}~\bibnamefont {Mesot}}, \bibinfo {author} {\bibfnamefont
	  {Ruixing}\ \bibnamefont {Liang}}, \bibinfo {author} {\bibfnamefont
	  {DA}~\bibnamefont {Bonn}}, \bibinfo {author} {\bibfnamefont {WN}~\bibnamefont
	  {Hardy}}, \bibinfo {author} {\bibfnamefont {A}~\bibnamefont {Watenphul}},
	  \emph {et~al.},\ }\bibfield  {title} {\enquote {\bibinfo {title} {Direct
	  observation of competition between superconductivity and charge density wave
	  order in {YBa$_2$Cu$_3$O$_{6.67}$}},}\ }\href@noop {} {\bibfield  {journal}
	  {\bibinfo  {journal} {Nature Physics}\ }\textbf {\bibinfo {volume} {8}},\
	  \bibinfo {pages} {871--876} (\bibinfo {year} {2012})}\BibitemShut {NoStop}%
	\bibitem [{\citenamefont {Croft}\ \emph {et~al.}(2014)\citenamefont {Croft},
	  \citenamefont {Lester}, \citenamefont {Senn}, \citenamefont {Bombardi},\ and\
	  \citenamefont {Hayden}}]{croft2014charge}%
	  \BibitemOpen
	  \bibfield  {author} {\bibinfo {author} {\bibfnamefont {TP}~\bibnamefont
	  {Croft}}, \bibinfo {author} {\bibfnamefont {Christopher}\ \bibnamefont
	  {Lester}}, \bibinfo {author} {\bibfnamefont {MS}~\bibnamefont {Senn}},
	  \bibinfo {author} {\bibfnamefont {Alessandro}\ \bibnamefont {Bombardi}}, \
	  and\ \bibinfo {author} {\bibfnamefont {SM}~\bibnamefont {Hayden}},\
	  }\bibfield  {title} {\enquote {\bibinfo {title} {Charge density wave
	  fluctuations in {La$_{2-x}$Sr$_x$CuO$_4$} and their competition with
	  superconductivity},}\ }\href@noop {} {\bibfield  {journal} {\bibinfo
	  {journal} {Physical Review B}\ }\textbf {\bibinfo {volume} {89}},\ \bibinfo
	  {pages} {224513} (\bibinfo {year} {2014})}\BibitemShut {NoStop}%
	\bibitem [{\citenamefont {Comin}\ \emph {et~al.}(2015)\citenamefont {Comin},
	  \citenamefont {Sutarto}, \citenamefont {He}, \citenamefont {da~Silva~Neto},
	  \citenamefont {Chauviere}, \citenamefont {Frano}, \citenamefont {Liang},
	  \citenamefont {Hardy}, \citenamefont {Bonn}, \citenamefont {Yoshida} \emph
	  {et~al.}}]{comin2015symmetry}%
	  \BibitemOpen
	  \bibfield  {author} {\bibinfo {author} {\bibfnamefont {R}~\bibnamefont
	  {Comin}}, \bibinfo {author} {\bibfnamefont {R}~\bibnamefont {Sutarto}},
	  \bibinfo {author} {\bibfnamefont {F}~\bibnamefont {He}}, \bibinfo {author}
	  {\bibfnamefont {EH}~\bibnamefont {da~Silva~Neto}}, \bibinfo {author}
	  {\bibfnamefont {L}~\bibnamefont {Chauviere}}, \bibinfo {author}
	  {\bibfnamefont {A}~\bibnamefont {Frano}}, \bibinfo {author} {\bibfnamefont
	  {R}~\bibnamefont {Liang}}, \bibinfo {author} {\bibfnamefont {WN}~\bibnamefont
	  {Hardy}}, \bibinfo {author} {\bibfnamefont {DA}~\bibnamefont {Bonn}},
	  \bibinfo {author} {\bibfnamefont {Y}~\bibnamefont {Yoshida}},  \emph
	  {et~al.},\ }\bibfield  {title} {\enquote {\bibinfo {title} {Symmetry of
	  charge order in cuprates},}\ }\href@noop {} {\bibfield  {journal} {\bibinfo
	  {journal} {Nature materials}\ }\textbf {\bibinfo {volume} {14}},\ \bibinfo
	  {pages} {796--800} (\bibinfo {year} {2015})}\BibitemShut {NoStop}%
	\bibitem [{\citenamefont {Comin}\ \emph {et~al.}(2014)\citenamefont {Comin},
	  \citenamefont {Frano}, \citenamefont {Yee}, \citenamefont {Yoshida},
	  \citenamefont {Eisaki}, \citenamefont {Schierle}, \citenamefont {Weschke},
	  \citenamefont {Sutarto}, \citenamefont {He}, \citenamefont {Soumyanarayanan}
	  \emph {et~al.}}]{comin2014charge}%
	  \BibitemOpen
	  \bibfield  {author} {\bibinfo {author} {\bibfnamefont {R}~\bibnamefont
	  {Comin}}, \bibinfo {author} {\bibfnamefont {A}~\bibnamefont {Frano}},
	  \bibinfo {author} {\bibfnamefont {Michael~Manchun}\ \bibnamefont {Yee}},
	  \bibinfo {author} {\bibfnamefont {Y}~\bibnamefont {Yoshida}}, \bibinfo
	  {author} {\bibfnamefont {H}~\bibnamefont {Eisaki}}, \bibinfo {author}
	  {\bibfnamefont {E}~\bibnamefont {Schierle}}, \bibinfo {author} {\bibfnamefont
	  {E}~\bibnamefont {Weschke}}, \bibinfo {author} {\bibfnamefont
	  {R}~\bibnamefont {Sutarto}}, \bibinfo {author} {\bibfnamefont
	  {F}~\bibnamefont {He}}, \bibinfo {author} {\bibfnamefont {Anjan}\
	  \bibnamefont {Soumyanarayanan}},  \emph {et~al.},\ }\bibfield  {title}
	  {\enquote {\bibinfo {title} {Charge order driven by fermi-arc instability in
	  {Bi$_2$Sr$_{2-x}$La$_x$CuO$_{6+\delta}$}},}\ }\href@noop {} {\bibfield
	  {journal} {\bibinfo  {journal} {Science}\ }\textbf {\bibinfo {volume}
	  {343}},\ \bibinfo {pages} {390--392} (\bibinfo {year} {2014})}\BibitemShut
	  {NoStop}%
	\bibitem [{\citenamefont {Blackburn}\ \emph {et~al.}(2013)\citenamefont
	  {Blackburn}, \citenamefont {Chang}, \citenamefont {H{\"u}cker}, \citenamefont
	  {Holmes}, \citenamefont {Christensen}, \citenamefont {Liang}, \citenamefont
	  {Bonn}, \citenamefont {Hardy}, \citenamefont {R{\"u}tt}, \citenamefont
	  {Gutowski} \emph {et~al.}}]{blackburn2013x}%
	  \BibitemOpen
	  \bibfield  {author} {\bibinfo {author} {\bibfnamefont {E}~\bibnamefont
	  {Blackburn}}, \bibinfo {author} {\bibfnamefont {J}~\bibnamefont {Chang}},
	  \bibinfo {author} {\bibfnamefont {M}~\bibnamefont {H{\"u}cker}}, \bibinfo
	  {author} {\bibfnamefont {AT}~\bibnamefont {Holmes}}, \bibinfo {author}
	  {\bibfnamefont {Niels~Bech}\ \bibnamefont {Christensen}}, \bibinfo {author}
	  {\bibfnamefont {Ruixing}\ \bibnamefont {Liang}}, \bibinfo {author}
	  {\bibfnamefont {DA}~\bibnamefont {Bonn}}, \bibinfo {author} {\bibfnamefont
	  {WN}~\bibnamefont {Hardy}}, \bibinfo {author} {\bibfnamefont {U}~\bibnamefont
	  {R{\"u}tt}}, \bibinfo {author} {\bibfnamefont {Olof}\ \bibnamefont
	  {Gutowski}},  \emph {et~al.},\ }\bibfield  {title} {\enquote {\bibinfo
	  {title} {X-ray diffraction observations of a charge-density-wave order in
	  superconducting ortho-{II} {YBa$_2$Cu$_3$O$_{6.54}$} single crystals in zero
	  magnetic field},}\ }\href@noop {} {\bibfield  {journal} {\bibinfo  {journal}
	  {Physical review letters}\ }\textbf {\bibinfo {volume} {110}},\ \bibinfo
	  {pages} {137004} (\bibinfo {year} {2013})}\BibitemShut {NoStop}%
	\bibitem [{\citenamefont {Hamidian}\ \emph {et~al.}(2016)\citenamefont
	  {Hamidian}, \citenamefont {Edkins}, \citenamefont {Joo}, \citenamefont
	  {Kostin}, \citenamefont {Eisaki}, \citenamefont {Uchida}, \citenamefont
	  {Lawler}, \citenamefont {Kim}, \citenamefont {Mackenzie}, \citenamefont
	  {Fujita} \emph {et~al.}}]{hamidian2016detection}%
	  \BibitemOpen
	  \bibfield  {author} {\bibinfo {author} {\bibfnamefont {MH}~\bibnamefont
	  {Hamidian}}, \bibinfo {author} {\bibfnamefont {SD}~\bibnamefont {Edkins}},
	  \bibinfo {author} {\bibfnamefont {Sang~Hyun}\ \bibnamefont {Joo}}, \bibinfo
	  {author} {\bibfnamefont {A}~\bibnamefont {Kostin}}, \bibinfo {author}
	  {\bibfnamefont {H}~\bibnamefont {Eisaki}}, \bibinfo {author} {\bibfnamefont
	  {S}~\bibnamefont {Uchida}}, \bibinfo {author} {\bibfnamefont
	  {MJ}~\bibnamefont {Lawler}}, \bibinfo {author} {\bibfnamefont {E-A}\
	  \bibnamefont {Kim}}, \bibinfo {author} {\bibfnamefont {AP}~\bibnamefont
	  {Mackenzie}}, \bibinfo {author} {\bibfnamefont {K}~\bibnamefont {Fujita}},
	  \emph {et~al.},\ }\bibfield  {title} {\enquote {\bibinfo {title} {Detection
	  of a {C}ooper-pair density wave in {Bi$_2$Sr$_2$CaCu$_2$O$_{8+ x}$}},}\
	  }\href@noop {} {\bibfield  {journal} {\bibinfo  {journal} {Nature}\ }\textbf
	  {\bibinfo {volume} {532}},\ \bibinfo {pages} {343--347} (\bibinfo {year}
	  {2016})}\BibitemShut {NoStop}%
	\bibitem [{\citenamefont {Ruan}\ \emph {et~al.}(2018)\citenamefont {Ruan},
	  \citenamefont {Li}, \citenamefont {Hu}, \citenamefont {Hao}, \citenamefont
	  {Li}, \citenamefont {Cai}, \citenamefont {Zhou}, \citenamefont {Lee},\ and\
	  \citenamefont {Wang}}]{ruan2018visualization}%
	  \BibitemOpen
	  \bibfield  {author} {\bibinfo {author} {\bibfnamefont {Wei}\ \bibnamefont
	  {Ruan}}, \bibinfo {author} {\bibfnamefont {Xintong}\ \bibnamefont {Li}},
	  \bibinfo {author} {\bibfnamefont {Cheng}\ \bibnamefont {Hu}}, \bibinfo
	  {author} {\bibfnamefont {Zhenqi}\ \bibnamefont {Hao}}, \bibinfo {author}
	  {\bibfnamefont {Haiwei}\ \bibnamefont {Li}}, \bibinfo {author} {\bibfnamefont
	  {Peng}\ \bibnamefont {Cai}}, \bibinfo {author} {\bibfnamefont {Xingjiang}\
	  \bibnamefont {Zhou}}, \bibinfo {author} {\bibfnamefont {Dung-Hai}\
	  \bibnamefont {Lee}}, \ and\ \bibinfo {author} {\bibfnamefont {Yayu}\
	  \bibnamefont {Wang}},\ }\bibfield  {title} {\enquote {\bibinfo {title}
	  {Visualization of the periodic modulation of {C}ooper pairing in a cuprate
	  superconductor},}\ }\href@noop {} {\bibfield  {journal} {\bibinfo  {journal}
	  {Nature Physics}\ }\textbf {\bibinfo {volume} {14}},\ \bibinfo {pages}
	  {1178--1182} (\bibinfo {year} {2018})}\BibitemShut {NoStop}%
	\bibitem [{\citenamefont {Atkinson}\ \emph {et~al.}(2015)\citenamefont
	  {Atkinson}, \citenamefont {Kampf},\ and\ \citenamefont
	  {Bulut}}]{atkinson2015charge}%
	  \BibitemOpen
	  \bibfield  {author} {\bibinfo {author} {\bibfnamefont {WA}~\bibnamefont
	  {Atkinson}}, \bibinfo {author} {\bibfnamefont {Arno~P}\ \bibnamefont
	  {Kampf}}, \ and\ \bibinfo {author} {\bibfnamefont {S}~\bibnamefont {Bulut}},\
	  }\bibfield  {title} {\enquote {\bibinfo {title} {Charge order in the
	  pseudogap phase of cuprate superconductors},}\ }\href@noop {} {\bibfield
	  {journal} {\bibinfo  {journal} {New Journal of Physics}\ }\textbf {\bibinfo
	  {volume} {17}},\ \bibinfo {pages} {013025} (\bibinfo {year}
	  {2015})}\BibitemShut {NoStop}%
	\bibitem [{\citenamefont {Badoux}\ \emph {et~al.}(2016)\citenamefont {Badoux},
	  \citenamefont {Tabis}, \citenamefont {Lalibert{\'e}}, \citenamefont
	  {Grissonnanche}, \citenamefont {Vignolle}, \citenamefont {Vignolles},
	  \citenamefont {B{\'e}ard}, \citenamefont {Bonn}, \citenamefont {Hardy},
	  \citenamefont {Liang} \emph {et~al.}}]{badoux2016change}%
	  \BibitemOpen
	  \bibfield  {author} {\bibinfo {author} {\bibfnamefont {S}~\bibnamefont
	  {Badoux}}, \bibinfo {author} {\bibfnamefont {W}~\bibnamefont {Tabis}},
	  \bibinfo {author} {\bibfnamefont {F}~\bibnamefont {Lalibert{\'e}}}, \bibinfo
	  {author} {\bibfnamefont {G}~\bibnamefont {Grissonnanche}}, \bibinfo {author}
	  {\bibfnamefont {B}~\bibnamefont {Vignolle}}, \bibinfo {author} {\bibfnamefont
	  {D}~\bibnamefont {Vignolles}}, \bibinfo {author} {\bibfnamefont {Jerome}\
	  \bibnamefont {B{\'e}ard}}, \bibinfo {author} {\bibfnamefont {DA}~\bibnamefont
	  {Bonn}}, \bibinfo {author} {\bibfnamefont {WN}~\bibnamefont {Hardy}},
	  \bibinfo {author} {\bibfnamefont {R}~\bibnamefont {Liang}},  \emph {et~al.},\
	  }\bibfield  {title} {\enquote {\bibinfo {title} {Change of carrier density at
	  the pseudogap critical point of a cuprate superconductor},}\ }\href@noop {}
	  {\bibfield  {journal} {\bibinfo  {journal} {Nature}\ }\textbf {\bibinfo
	  {volume} {531}},\ \bibinfo {pages} {210--214} (\bibinfo {year}
	  {2016})}\BibitemShut {NoStop}%
	\bibitem [{\citenamefont {Verret}\ \emph {et~al.}(2017)\citenamefont {Verret},
	  \citenamefont {Charlebois}, \citenamefont {S{\'e}n{\'e}chal},\ and\
	  \citenamefont {Tremblay}}]{verret2017subgap}%
	  \BibitemOpen
	  \bibfield  {author} {\bibinfo {author} {\bibfnamefont {S}~\bibnamefont
	  {Verret}}, \bibinfo {author} {\bibfnamefont {M}~\bibnamefont {Charlebois}},
	  \bibinfo {author} {\bibfnamefont {D}~\bibnamefont {S{\'e}n{\'e}chal}}, \ and\
	  \bibinfo {author} {\bibfnamefont {A-MS}\ \bibnamefont {Tremblay}},\
	  }\bibfield  {title} {\enquote {\bibinfo {title} {Subgap structures and
	  pseudogap in cuprate superconductors: Role of density waves},}\ }\href@noop
	  {} {\bibfield  {journal} {\bibinfo  {journal} {Physical Review B}\ }\textbf
	  {\bibinfo {volume} {95}},\ \bibinfo {pages} {054518} (\bibinfo {year}
	  {2017})}\BibitemShut {NoStop}%
	\bibitem [{\citenamefont {Davis}\ and\ \citenamefont
	  {Lee}(2013)}]{davis2013concepts}%
	  \BibitemOpen
	  \bibfield  {author} {\bibinfo {author} {\bibfnamefont {JC~S{\'e}amus}\
	  \bibnamefont {Davis}}\ and\ \bibinfo {author} {\bibfnamefont {Dung-Hai}\
	  \bibnamefont {Lee}},\ }\bibfield  {title} {\enquote {\bibinfo {title}
	  {Concepts relating magnetic interactions, intertwined electronic orders, and
	  strongly correlated superconductivity},}\ }\href@noop {} {\bibfield
	  {journal} {\bibinfo  {journal} {Proceedings of the National Academy of
	  Sciences}\ }\textbf {\bibinfo {volume} {110}},\ \bibinfo {pages}
	  {17623--17630} (\bibinfo {year} {2013})}\BibitemShut {NoStop}%
	\bibitem [{\citenamefont {Sau}\ and\ \citenamefont
	  {Sachdev}(2014)}]{sau2014mean}%
	  \BibitemOpen
	  \bibfield  {author} {\bibinfo {author} {\bibfnamefont {Jay~D}\ \bibnamefont
	  {Sau}}\ and\ \bibinfo {author} {\bibfnamefont {Subir}\ \bibnamefont
	  {Sachdev}},\ }\bibfield  {title} {\enquote {\bibinfo {title} {Mean-field
	  theory of competing orders in metals with antiferromagnetic exchange
	  interactions},}\ }\href@noop {} {\bibfield  {journal} {\bibinfo  {journal}
	  {Physical Review B}\ }\textbf {\bibinfo {volume} {89}},\ \bibinfo {pages}
	  {075129} (\bibinfo {year} {2014})}\BibitemShut {NoStop}%
	\bibitem [{\citenamefont {Sachdev}\ and\ \citenamefont
	  {La~Placa}(2013)}]{sachdev2013bond}%
	  \BibitemOpen
	  \bibfield  {author} {\bibinfo {author} {\bibfnamefont {Subir}\ \bibnamefont
	  {Sachdev}}\ and\ \bibinfo {author} {\bibfnamefont {Rolando}\ \bibnamefont
	  {La~Placa}},\ }\bibfield  {title} {\enquote {\bibinfo {title} {Bond order in
	  two-dimensional metals with antiferromagnetic exchange interactions},}\
	  }\href@noop {} {\bibfield  {journal} {\bibinfo  {journal} {Physical review
	  letters}\ }\textbf {\bibinfo {volume} {111}},\ \bibinfo {pages} {027202}
	  (\bibinfo {year} {2013})}\BibitemShut {NoStop}%
	\bibitem [{\citenamefont {Vojta}\ \emph {et~al.}(2000)\citenamefont {Vojta},
	  \citenamefont {Zhang},\ and\ \citenamefont {Sachdev}}]{vojta2000competing}%
	  \BibitemOpen
	  \bibfield  {author} {\bibinfo {author} {\bibfnamefont {Matthias}\
	  \bibnamefont {Vojta}}, \bibinfo {author} {\bibfnamefont {Ying}\ \bibnamefont
	  {Zhang}}, \ and\ \bibinfo {author} {\bibfnamefont {Subir}\ \bibnamefont
	  {Sachdev}},\ }\bibfield  {title} {\enquote {\bibinfo {title} {Competing
	  orders and quantum criticality in doped antiferromagnets},}\ }\href@noop {}
	  {\bibfield  {journal} {\bibinfo  {journal} {Physical Review B}\ }\textbf
	  {\bibinfo {volume} {62}},\ \bibinfo {pages} {6721} (\bibinfo {year}
	  {2000})}\BibitemShut {NoStop}%
	\bibitem [{\citenamefont {Vojta}(2002)}]{vojta2002superconducting}%
	  \BibitemOpen
	  \bibfield  {author} {\bibinfo {author} {\bibfnamefont {Matthias}\
	  \bibnamefont {Vojta}},\ }\bibfield  {title} {\enquote {\bibinfo {title}
	  {Superconducting charge-ordered states in cuprates},}\ }\href@noop {}
	  {\bibfield  {journal} {\bibinfo  {journal} {Physical Review B}\ }\textbf
	  {\bibinfo {volume} {66}},\ \bibinfo {pages} {104505} (\bibinfo {year}
	  {2002})}\BibitemShut {NoStop}%
	\bibitem [{\citenamefont {Bejas}\ \emph {et~al.}(2012)\citenamefont {Bejas},
	  \citenamefont {Greco},\ and\ \citenamefont {Yamase}}]{bejas2012possible}%
	  \BibitemOpen
	  \bibfield  {author} {\bibinfo {author} {\bibfnamefont {Matias}\ \bibnamefont
	  {Bejas}}, \bibinfo {author} {\bibfnamefont {Andr{\'e}s}\ \bibnamefont
	  {Greco}}, \ and\ \bibinfo {author} {\bibfnamefont {Hiroyuki}\ \bibnamefont
	  {Yamase}},\ }\bibfield  {title} {\enquote {\bibinfo {title} {Possible charge
	  instabilities in two-dimensional doped mott insulators},}\ }\href@noop {}
	  {\bibfield  {journal} {\bibinfo  {journal} {Physical Review B}\ }\textbf
	  {\bibinfo {volume} {86}},\ \bibinfo {pages} {224509} (\bibinfo {year}
	  {2012})}\BibitemShut {NoStop}%
	\bibitem [{\citenamefont {Raczkowski}\ \emph {et~al.}(2007)\citenamefont
	  {Raczkowski}, \citenamefont {Capello}, \citenamefont {Poilblanc},
	  \citenamefont {Fr{\'e}sard},\ and\ \citenamefont
	  {Ole{\'s}}}]{raczkowski2007unidirectional}%
	  \BibitemOpen
	  \bibfield  {author} {\bibinfo {author} {\bibfnamefont {Marcin}\ \bibnamefont
	  {Raczkowski}}, \bibinfo {author} {\bibfnamefont {Manuela}\ \bibnamefont
	  {Capello}}, \bibinfo {author} {\bibfnamefont {Didier}\ \bibnamefont
	  {Poilblanc}}, \bibinfo {author} {\bibfnamefont {Raymond}\ \bibnamefont
	  {Fr{\'e}sard}}, \ and\ \bibinfo {author} {\bibfnamefont {Andrzej~M}\
	  \bibnamefont {Ole{\'s}}},\ }\bibfield  {title} {\enquote {\bibinfo {title}
	  {Unidirectional d-wave superconducting domains in the two-dimensional
	  $t$-${J}$ model},}\ }\href@noop {} {\bibfield  {journal} {\bibinfo  {journal}
	  {Physical Review B}\ }\textbf {\bibinfo {volume} {76}},\ \bibinfo {pages}
	  {140505} (\bibinfo {year} {2007})}\BibitemShut {NoStop}%
	\bibitem [{\citenamefont {Allais}\ \emph
	  {et~al.}(2014{\natexlab{a}})\citenamefont {Allais}, \citenamefont {Bauer},\
	  and\ \citenamefont {Sachdev}}]{allais2014density}%
	  \BibitemOpen
	  \bibfield  {author} {\bibinfo {author} {\bibfnamefont {Andrea}\ \bibnamefont
	  {Allais}}, \bibinfo {author} {\bibfnamefont {Johannes}\ \bibnamefont
	  {Bauer}}, \ and\ \bibinfo {author} {\bibfnamefont {Subir}\ \bibnamefont
	  {Sachdev}},\ }\bibfield  {title} {\enquote {\bibinfo {title} {Density wave
	  instabilities in a correlated two-dimensional metal},}\ }\href@noop {}
	  {\bibfield  {journal} {\bibinfo  {journal} {Physical Review B}\ }\textbf
	  {\bibinfo {volume} {90}},\ \bibinfo {pages} {155114} (\bibinfo {year}
	  {2014}{\natexlab{a}})}\BibitemShut {NoStop}%
	\bibitem [{\citenamefont {P{\'e}pin}\ \emph {et~al.}(2014)\citenamefont
	  {P{\'e}pin}, \citenamefont {De~Carvalho}, \citenamefont {Kloss},\ and\
	  \citenamefont {Montiel}}]{pepin2014pseudogap}%
	  \BibitemOpen
	  \bibfield  {author} {\bibinfo {author} {\bibfnamefont {C}~\bibnamefont
	  {P{\'e}pin}}, \bibinfo {author} {\bibfnamefont {VS}~\bibnamefont
	  {De~Carvalho}}, \bibinfo {author} {\bibfnamefont {T}~\bibnamefont {Kloss}}, \
	  and\ \bibinfo {author} {\bibfnamefont {X}~\bibnamefont {Montiel}},\
	  }\bibfield  {title} {\enquote {\bibinfo {title} {Pseudogap, charge order, and
	  pairing density wave at the hot spots in cuprate superconductors},}\
	  }\href@noop {} {\bibfield  {journal} {\bibinfo  {journal} {Physical Review
	  B}\ }\textbf {\bibinfo {volume} {90}},\ \bibinfo {pages} {195207} (\bibinfo
	  {year} {2014})}\BibitemShut {NoStop}%
	\bibitem [{\citenamefont {Wang}\ and\ \citenamefont
	  {Chubukov}(2014)}]{wang2014charge}%
	  \BibitemOpen
	  \bibfield  {author} {\bibinfo {author} {\bibfnamefont {Yuxuan}\ \bibnamefont
	  {Wang}}\ and\ \bibinfo {author} {\bibfnamefont {Andrey}\ \bibnamefont
	  {Chubukov}},\ }\bibfield  {title} {\enquote {\bibinfo {title}
	  {Charge-density-wave order with momentum {$(2\textbf{Q}, 0)$} and
	  {$(0,2\textbf{Q})$} within the spin-fermion model: Continuous and discrete
	  symmetry breaking, preemptive composite order, and relation to pseudogap in
	  hole-doped cuprates},}\ }\href@noop {} {\bibfield  {journal} {\bibinfo
	  {journal} {Physical Review B}\ }\textbf {\bibinfo {volume} {90}},\ \bibinfo
	  {pages} {035149} (\bibinfo {year} {2014})}\BibitemShut {NoStop}%
	\bibitem [{\citenamefont {Meier}\ \emph {et~al.}(2014)\citenamefont {Meier},
	  \citenamefont {P{\'e}pin}, \citenamefont {Einenkel},\ and\ \citenamefont
	  {Efetov}}]{meier2014cascade}%
	  \BibitemOpen
	  \bibfield  {author} {\bibinfo {author} {\bibfnamefont {H}~\bibnamefont
	  {Meier}}, \bibinfo {author} {\bibfnamefont {C}~\bibnamefont {P{\'e}pin}},
	  \bibinfo {author} {\bibfnamefont {M}~\bibnamefont {Einenkel}}, \ and\
	  \bibinfo {author} {\bibfnamefont {KB}~\bibnamefont {Efetov}},\ }\bibfield
	  {title} {\enquote {\bibinfo {title} {Cascade of phase transitions in the
	  vicinity of a quantum critical point},}\ }\href@noop {} {\bibfield  {journal}
	  {\bibinfo  {journal} {Physical Review B}\ }\textbf {\bibinfo {volume} {89}},\
	  \bibinfo {pages} {195115} (\bibinfo {year} {2014})}\BibitemShut {NoStop}%
	\bibitem [{\citenamefont {Pietig}\ \emph {et~al.}(1999)\citenamefont {Pietig},
	  \citenamefont {Bulla},\ and\ \citenamefont {Blawid}}]{pietig1999reentrant}%
	  \BibitemOpen
	  \bibfield  {author} {\bibinfo {author} {\bibfnamefont {R}~\bibnamefont
	  {Pietig}}, \bibinfo {author} {\bibfnamefont {R}~\bibnamefont {Bulla}}, \ and\
	  \bibinfo {author} {\bibfnamefont {S}~\bibnamefont {Blawid}},\ }\bibfield
	  {title} {\enquote {\bibinfo {title} {Reentrant charge order transition in the
	  extended hubbard model},}\ }\href@noop {} {\bibfield  {journal} {\bibinfo
	  {journal} {Physical review letters}\ }\textbf {\bibinfo {volume} {82}},\
	  \bibinfo {pages} {4046} (\bibinfo {year} {1999})}\BibitemShut {NoStop}%
	\bibitem [{\citenamefont {Tong}\ \emph {et~al.}(2004)\citenamefont {Tong},
	  \citenamefont {Shen},\ and\ \citenamefont {Bulla}}]{tong2004charge}%
	  \BibitemOpen
	  \bibfield  {author} {\bibinfo {author} {\bibfnamefont {Ning-Hua}\
	  \bibnamefont {Tong}}, \bibinfo {author} {\bibfnamefont {Shun-Qing}\
	  \bibnamefont {Shen}}, \ and\ \bibinfo {author} {\bibfnamefont {Ralf}\
	  \bibnamefont {Bulla}},\ }\bibfield  {title} {\enquote {\bibinfo {title}
	  {Charge ordering and phase separation in the infinite dimensional extended
	  hubbard model},}\ }\href@noop {} {\bibfield  {journal} {\bibinfo  {journal}
	  {Physical Review B}\ }\textbf {\bibinfo {volume} {70}},\ \bibinfo {pages}
	  {085118} (\bibinfo {year} {2004})}\BibitemShut {NoStop}%
	\bibitem [{\citenamefont {Amaricci}\ \emph {et~al.}(2010)\citenamefont
	  {Amaricci}, \citenamefont {Camjayi}, \citenamefont {Haule}, \citenamefont
	  {Kotliar}, \citenamefont {Tanaskovi{\'c}},\ and\ \citenamefont
	  {Dobrosavljevi{\'c}}}]{amaricci2010extended}%
	  \BibitemOpen
	  \bibfield  {author} {\bibinfo {author} {\bibfnamefont {Adriano}\ \bibnamefont
	  {Amaricci}}, \bibinfo {author} {\bibfnamefont {Alberto}\ \bibnamefont
	  {Camjayi}}, \bibinfo {author} {\bibfnamefont {Kristjan}\ \bibnamefont
	  {Haule}}, \bibinfo {author} {\bibfnamefont {G}~\bibnamefont {Kotliar}},
	  \bibinfo {author} {\bibfnamefont {D}~\bibnamefont {Tanaskovi{\'c}}}, \ and\
	  \bibinfo {author} {\bibfnamefont {V}~\bibnamefont {Dobrosavljevi{\'c}}},\
	  }\bibfield  {title} {\enquote {\bibinfo {title} {Extended hubbard model:
	  Charge ordering and wigner-mott transition},}\ }\href@noop {} {\bibfield
	  {journal} {\bibinfo  {journal} {Physical Review B}\ }\textbf {\bibinfo
	  {volume} {82}},\ \bibinfo {pages} {155102} (\bibinfo {year}
	  {2010})}\BibitemShut {NoStop}%
	\bibitem [{\citenamefont {Allais}\ \emph
	  {et~al.}(2014{\natexlab{b}})\citenamefont {Allais}, \citenamefont {Bauer},\
	  and\ \citenamefont {Sachdev}}]{allais2014auxiliary}%
	  \BibitemOpen
	  \bibfield  {author} {\bibinfo {author} {\bibfnamefont {A}~\bibnamefont
	  {Allais}}, \bibinfo {author} {\bibfnamefont {J}~\bibnamefont {Bauer}}, \ and\
	  \bibinfo {author} {\bibfnamefont {S}~\bibnamefont {Sachdev}},\ }\bibfield
	  {title} {\enquote {\bibinfo {title} {Auxiliary-boson and dmft studies of bond
	  ordering instabilities of formula not shown-formula not shown-formula not
	  shown models on the square lattice},}\ }\href@noop {} {\bibfield  {journal}
	  {\bibinfo  {journal} {INDIAN JOURNAL OF PHYSICS}\ }\textbf {\bibinfo {volume}
	  {88}},\ \bibinfo {pages} {905--913} (\bibinfo {year}
	  {2014}{\natexlab{b}})}\BibitemShut {NoStop}%
	\bibitem [{\citenamefont {Terletska}\ \emph {et~al.}(2017)\citenamefont
	  {Terletska}, \citenamefont {Chen},\ and\ \citenamefont
	  {Gull}}]{terletska2017charge}%
	  \BibitemOpen
	  \bibfield  {author} {\bibinfo {author} {\bibfnamefont {Hanna}\ \bibnamefont
	  {Terletska}}, \bibinfo {author} {\bibfnamefont {Tianran}\ \bibnamefont
	  {Chen}}, \ and\ \bibinfo {author} {\bibfnamefont {Emanuel}\ \bibnamefont
	  {Gull}},\ }\bibfield  {title} {\enquote {\bibinfo {title} {Charge ordering
	  and correlation effects in the extended {H}ubbard model},}\ }\href@noop {}
	  {\bibfield  {journal} {\bibinfo  {journal} {Physical Review B}\ }\textbf
	  {\bibinfo {volume} {95}},\ \bibinfo {pages} {115149} (\bibinfo {year}
	  {2017})}\BibitemShut {NoStop}%
	\bibitem [{\citenamefont {Faye}\ and\ \citenamefont
	  {S{\'e}n{\'e}chal}(2017)}]{faye2017interplay}%
	  \BibitemOpen
	  \bibfield  {author} {\bibinfo {author} {\bibfnamefont {JPL}\ \bibnamefont
	  {Faye}}\ and\ \bibinfo {author} {\bibfnamefont {D}~\bibnamefont
	  {S{\'e}n{\'e}chal}},\ }\bibfield  {title} {\enquote {\bibinfo {title}
	  {Interplay between $d$-wave superconductivity and a bond-density wave in the
	  one-band hubbard model},}\ }\href@noop {} {\bibfield  {journal} {\bibinfo
	  {journal} {Physical Review B}\ }\textbf {\bibinfo {volume} {95}},\ \bibinfo
	  {pages} {115127} (\bibinfo {year} {2017})}\BibitemShut {NoStop}%
	\bibitem [{\citenamefont {Freire}\ \emph {et~al.}(2015)\citenamefont {Freire},
	  \citenamefont {De~Carvalho},\ and\ \citenamefont
	  {P{\'e}pin}}]{freire2015renormalization}%
	  \BibitemOpen
	  \bibfield  {author} {\bibinfo {author} {\bibfnamefont {Hermann}\ \bibnamefont
	  {Freire}}, \bibinfo {author} {\bibfnamefont {Vanuildo~S}\ \bibnamefont
	  {De~Carvalho}}, \ and\ \bibinfo {author} {\bibfnamefont {Catherine}\
	  \bibnamefont {P{\'e}pin}},\ }\bibfield  {title} {\enquote {\bibinfo {title}
	  {Renormalization group analysis of the pair-density-wave and charge order
	  within the fermionic hot-spot model for cuprate superconductors},}\
	  }\href@noop {} {\bibfield  {journal} {\bibinfo  {journal} {Physical Review
	  B}\ }\textbf {\bibinfo {volume} {92}},\ \bibinfo {pages} {045132} (\bibinfo
	  {year} {2015})}\BibitemShut {NoStop}%
	\bibitem [{\citenamefont {White}\ and\ \citenamefont
	  {Scalapino}(1999)}]{white1999competition}%
	  \BibitemOpen
	  \bibfield  {author} {\bibinfo {author} {\bibfnamefont {Steven~R}\
	  \bibnamefont {White}}\ and\ \bibinfo {author} {\bibfnamefont
	  {DJ}~\bibnamefont {Scalapino}},\ }\bibfield  {title} {\enquote {\bibinfo
	  {title} {Competition between stripes and pairing in a $t$-$t'$-${J}$
	  model},}\ }\href@noop {} {\bibfield  {journal} {\bibinfo  {journal} {Physical
	  Review B}\ }\textbf {\bibinfo {volume} {60}},\ \bibinfo {pages} {R753}
	  (\bibinfo {year} {1999})}\BibitemShut {NoStop}%
	\bibitem [{\citenamefont {Bauer}\ and\ \citenamefont
	  {Hewson}(2010)}]{bauer2010competition}%
	  \BibitemOpen
	  \bibfield  {author} {\bibinfo {author} {\bibfnamefont {Johannes}\
	  \bibnamefont {Bauer}}\ and\ \bibinfo {author} {\bibfnamefont {Alex~C}\
	  \bibnamefont {Hewson}},\ }\bibfield  {title} {\enquote {\bibinfo {title}
	  {Competition between antiferromagnetic and charge order in the
	  hubbard-holstein model},}\ }\href@noop {} {\bibfield  {journal} {\bibinfo
	  {journal} {Physical Review B}\ }\textbf {\bibinfo {volume} {81}},\ \bibinfo
	  {pages} {235113} (\bibinfo {year} {2010})}\BibitemShut {NoStop}%
	\bibitem [{\citenamefont {Achkar}\ \emph {et~al.}(2012)\citenamefont {Achkar},
	  \citenamefont {Sutarto}, \citenamefont {Mao}, \citenamefont {He},
	  \citenamefont {Frano}, \citenamefont {Blanco-Canosa}, \citenamefont
	  {Le~Tacon}, \citenamefont {Ghiringhelli}, \citenamefont {Braicovich},
	  \citenamefont {Minola} \emph {et~al.}}]{achkar2012distinct}%
	  \BibitemOpen
	  \bibfield  {author} {\bibinfo {author} {\bibfnamefont {AJ}~\bibnamefont
	  {Achkar}}, \bibinfo {author} {\bibfnamefont {R}~\bibnamefont {Sutarto}},
	  \bibinfo {author} {\bibfnamefont {X}~\bibnamefont {Mao}}, \bibinfo {author}
	  {\bibfnamefont {F}~\bibnamefont {He}}, \bibinfo {author} {\bibfnamefont
	  {A}~\bibnamefont {Frano}}, \bibinfo {author} {\bibfnamefont {S}~\bibnamefont
	  {Blanco-Canosa}}, \bibinfo {author} {\bibfnamefont {M}~\bibnamefont
	  {Le~Tacon}}, \bibinfo {author} {\bibfnamefont {G}~\bibnamefont
	  {Ghiringhelli}}, \bibinfo {author} {\bibfnamefont {Lucio}\ \bibnamefont
	  {Braicovich}}, \bibinfo {author} {\bibfnamefont {Matteo}\ \bibnamefont
	  {Minola}},  \emph {et~al.},\ }\bibfield  {title} {\enquote {\bibinfo {title}
	  {Distinct charge orders in the planes and chains of ortho-iii-ordered
	  {YB}a$_2${C}u$_3${O}$_{6+\delta}$ superconductors identified by resonant
	  elastic x-ray scattering},}\ }\href@noop {} {\bibfield  {journal} {\bibinfo
	  {journal} {Physical review letters}\ }\textbf {\bibinfo {volume} {109}},\
	  \bibinfo {pages} {167001} (\bibinfo {year} {2012})}\BibitemShut {NoStop}%
	\bibitem [{\citenamefont {Blanco-Canosa}\ \emph {et~al.}(2013)\citenamefont
	  {Blanco-Canosa}, \citenamefont {Frano}, \citenamefont {Loew}, \citenamefont
	  {Lu}, \citenamefont {Porras}, \citenamefont {Ghiringhelli}, \citenamefont
	  {Minola}, \citenamefont {Mazzoli}, \citenamefont {Braicovich}, \citenamefont
	  {Schierle} \emph {et~al.}}]{blanco2013momentum}%
	  \BibitemOpen
	  \bibfield  {author} {\bibinfo {author} {\bibfnamefont {S}~\bibnamefont
	  {Blanco-Canosa}}, \bibinfo {author} {\bibfnamefont {A}~\bibnamefont {Frano}},
	  \bibinfo {author} {\bibfnamefont {T}~\bibnamefont {Loew}}, \bibinfo {author}
	  {\bibfnamefont {Y}~\bibnamefont {Lu}}, \bibinfo {author} {\bibfnamefont
	  {J}~\bibnamefont {Porras}}, \bibinfo {author} {\bibfnamefont {G}~\bibnamefont
	  {Ghiringhelli}}, \bibinfo {author} {\bibfnamefont {MATTEO}\ \bibnamefont
	  {Minola}}, \bibinfo {author} {\bibfnamefont {C}~\bibnamefont {Mazzoli}},
	  \bibinfo {author} {\bibfnamefont {LUCIO}\ \bibnamefont {Braicovich}},
	  \bibinfo {author} {\bibfnamefont {E}~\bibnamefont {Schierle}},  \emph
	  {et~al.},\ }\bibfield  {title} {\enquote {\bibinfo {title}
	  {Momentum-dependent charge correlations in {YB}a$_2${C}u$_3${O}$_{6+\delta}$
	  superconductors probed by resonant x-ray scattering: Evidence for three
	  competing phases},}\ }\href@noop {} {\bibfield  {journal} {\bibinfo
	  {journal} {Physical review letters}\ }\textbf {\bibinfo {volume} {110}},\
	  \bibinfo {pages} {187001} (\bibinfo {year} {2013})}\BibitemShut {NoStop}%
	\bibitem [{\citenamefont {Cappelluti}\ and\ \citenamefont
	  {Zeyher}(1999)}]{cappelluti1999interplay}%
	  \BibitemOpen
	  \bibfield  {author} {\bibinfo {author} {\bibfnamefont {Emmanuele}\
	  \bibnamefont {Cappelluti}}\ and\ \bibinfo {author} {\bibfnamefont
	  {R}~\bibnamefont {Zeyher}},\ }\bibfield  {title} {\enquote {\bibinfo {title}
	  {Interplay between superconductivity and flux phase in the $t$-${J}$
	  model},}\ }\href@noop {} {\bibfield  {journal} {\bibinfo  {journal} {Physical
	  Review B}\ }\textbf {\bibinfo {volume} {59}},\ \bibinfo {pages} {6475}
	  (\bibinfo {year} {1999})}\BibitemShut {NoStop}%
	\bibitem [{\citenamefont {Liechtenstein}\ \emph {et~al.}(1996)\citenamefont
	  {Liechtenstein}, \citenamefont {Gunnarsson}, \citenamefont {Andersen},\ and\
	  \citenamefont {Martin}}]{liechtenstein1996quasiparticle}%
	  \BibitemOpen
	  \bibfield  {author} {\bibinfo {author} {\bibfnamefont {AI}~\bibnamefont
	  {Liechtenstein}}, \bibinfo {author} {\bibfnamefont {Olle}\ \bibnamefont
	  {Gunnarsson}}, \bibinfo {author} {\bibfnamefont {OK}~\bibnamefont
	  {Andersen}}, \ and\ \bibinfo {author} {\bibfnamefont {RM}~\bibnamefont
	  {Martin}},\ }\bibfield  {title} {\enquote {\bibinfo {title} {Quasiparticle
	  bands and superconductivity in bilayer cuprates},}\ }\href@noop {} {\bibfield
	   {journal} {\bibinfo  {journal} {Physical Review B}\ }\textbf {\bibinfo
	  {volume} {54}},\ \bibinfo {pages} {12505} (\bibinfo {year}
	  {1996})}\BibitemShut {NoStop}%
	\bibitem [{\citenamefont {Andersen}\ \emph {et~al.}(1995)\citenamefont
	  {Andersen}, \citenamefont {Liechtenstein}, \citenamefont {Jepsen},\ and\
	  \citenamefont {Paulsen}}]{andersen1995lda}%
	  \BibitemOpen
	  \bibfield  {author} {\bibinfo {author} {\bibfnamefont {OK}~\bibnamefont
	  {Andersen}}, \bibinfo {author} {\bibfnamefont {AI}~\bibnamefont
	  {Liechtenstein}}, \bibinfo {author} {\bibfnamefont {O}~\bibnamefont
	  {Jepsen}}, \ and\ \bibinfo {author} {\bibfnamefont {F}~\bibnamefont
	  {Paulsen}},\ }\bibfield  {title} {\enquote {\bibinfo {title} {{LDA} energy
	  bands, low-energy hamiltonians, $t'$, $t''$, $t_{\perp}(k)$, and
	  \textit{{J}}$_{\perp}$},}\ }\href@noop {} {\bibfield  {journal} {\bibinfo
	  {journal} {Journal of Physics and Chemistry of Solids}\ }\textbf {\bibinfo
	  {volume} {56}},\ \bibinfo {pages} {1573--1591} (\bibinfo {year}
	  {1995})}\BibitemShut {NoStop}%
	\bibitem [{\citenamefont {Kotliar}\ \emph {et~al.}(2001)\citenamefont
	  {Kotliar}, \citenamefont {Savrasov}, \citenamefont {P{\'a}lsson},\ and\
	  \citenamefont {Biroli}}]{kotliar2001cellular}%
	  \BibitemOpen
	  \bibfield  {author} {\bibinfo {author} {\bibfnamefont {Gabriel}\ \bibnamefont
	  {Kotliar}}, \bibinfo {author} {\bibfnamefont {Sergej~Y}\ \bibnamefont
	  {Savrasov}}, \bibinfo {author} {\bibfnamefont {Gunnar}\ \bibnamefont
	  {P{\'a}lsson}}, \ and\ \bibinfo {author} {\bibfnamefont {Giulio}\
	  \bibnamefont {Biroli}},\ }\bibfield  {title} {\enquote {\bibinfo {title}
	  {Cellular dynamical mean field approach to strongly correlated systems},}\
	  }\href@noop {} {\bibfield  {journal} {\bibinfo  {journal} {Physical review
	  letters}\ }\textbf {\bibinfo {volume} {87}},\ \bibinfo {pages} {186401}
	  (\bibinfo {year} {2001})}\BibitemShut {NoStop}%
	\bibitem [{\citenamefont {Kancharla}\ \emph {et~al.}(2008)\citenamefont
	  {Kancharla}, \citenamefont {Kyung}, \citenamefont {S{\'e}n{\'e}chal},
	  \citenamefont {Civelli}, \citenamefont {Capone}, \citenamefont {Kotliar},\
	  and\ \citenamefont {Tremblay}}]{kancharla2008anomalous}%
	  \BibitemOpen
	  \bibfield  {author} {\bibinfo {author} {\bibfnamefont {SS}~\bibnamefont
	  {Kancharla}}, \bibinfo {author} {\bibfnamefont {B}~\bibnamefont {Kyung}},
	  \bibinfo {author} {\bibfnamefont {David}\ \bibnamefont {S{\'e}n{\'e}chal}},
	  \bibinfo {author} {\bibfnamefont {M}~\bibnamefont {Civelli}}, \bibinfo
	  {author} {\bibfnamefont {Massimo}\ \bibnamefont {Capone}}, \bibinfo {author}
	  {\bibfnamefont {G}~\bibnamefont {Kotliar}}, \ and\ \bibinfo {author}
	  {\bibfnamefont {A-MS}\ \bibnamefont {Tremblay}},\ }\bibfield  {title}
	  {\enquote {\bibinfo {title} {Anomalous superconductivity and its competition
	  with antiferromagnetism in doped {Mott} insulators},}\ }\href@noop {}
	  {\bibfield  {journal} {\bibinfo  {journal} {Physical Review B}\ }\textbf
	  {\bibinfo {volume} {77}},\ \bibinfo {pages} {184516} (\bibinfo {year}
	  {2008})}\BibitemShut {NoStop}%
	\bibitem [{\citenamefont {S{\'e}n{\'e}chal}(2010)}]{senechal2010bath}%
	  \BibitemOpen
	  \bibfield  {author} {\bibinfo {author} {\bibfnamefont {David}\ \bibnamefont
	  {S{\'e}n{\'e}chal}},\ }\bibfield  {title} {\enquote {\bibinfo {title} {Bath
	  optimization in the cellular dynamical mean field theory},}\ }\href@noop {}
	  {\bibfield  {journal} {\bibinfo  {journal} {Physical Review B}\ }\textbf
	  {\bibinfo {volume} {81}},\ \bibinfo {pages} {235125} (\bibinfo {year}
	  {2010})}\BibitemShut {NoStop}%
	\bibitem [{\citenamefont {S{\'e}n{\'e}chal}(2012)}]{senechal2012cluster}%
	  \BibitemOpen
	  \bibfield  {author} {\bibinfo {author} {\bibfnamefont {David}\ \bibnamefont
	  {S{\'e}n{\'e}chal}},\ }\bibfield  {title} {\enquote {\bibinfo {title}
	  {Cluster dynamical mean field theory},}\ }in\ \href@noop {} {\emph {\bibinfo
	  {booktitle} {Strongly Correlated Systems}}}\ (\bibinfo  {publisher}
	  {Springer},\ \bibinfo {year} {2012})\ pp.\ \bibinfo {pages}
	  {341--371}\BibitemShut {NoStop}%
	\bibitem [{\citenamefont {Koch}\ \emph {et~al.}(2008)\citenamefont {Koch},
	  \citenamefont {Sangiovanni},\ and\ \citenamefont {Gunnarsson}}]{koch2008sum}%
	  \BibitemOpen
	  \bibfield  {author} {\bibinfo {author} {\bibfnamefont {Erik}\ \bibnamefont
	  {Koch}}, \bibinfo {author} {\bibfnamefont {Giorgio}\ \bibnamefont
	  {Sangiovanni}}, \ and\ \bibinfo {author} {\bibfnamefont {Olle}\ \bibnamefont
	  {Gunnarsson}},\ }\bibfield  {title} {\enquote {\bibinfo {title} {Sum rules
	  and bath parametrization for quantum cluster theories},}\ }\href@noop {}
	  {\bibfield  {journal} {\bibinfo  {journal} {Physical Review B}\ }\textbf
	  {\bibinfo {volume} {78}},\ \bibinfo {pages} {115102} (\bibinfo {year}
	  {2008})}\BibitemShut {NoStop}%
	\bibitem [{\citenamefont {Foley}\ \emph {et~al.}(2019)\citenamefont {Foley},
	  \citenamefont {Verret}, \citenamefont {Tremblay},\ and\ \citenamefont
	  {Senechal}}]{foley2019coexistence}%
	  \BibitemOpen
	  \bibfield  {author} {\bibinfo {author} {\bibfnamefont {Alexandre}\
	  \bibnamefont {Foley}}, \bibinfo {author} {\bibfnamefont {Simon}\ \bibnamefont
	  {Verret}}, \bibinfo {author} {\bibfnamefont {A-MS}\ \bibnamefont {Tremblay}},
	  \ and\ \bibinfo {author} {\bibfnamefont {David}\ \bibnamefont {Senechal}},\
	  }\bibfield  {title} {\enquote {\bibinfo {title} {Coexistence of
	  superconductivity and antiferromagnetism in the {H}ubbard model for
	  cuprates},}\ }\href@noop {} {\bibfield  {journal} {\bibinfo  {journal}
	  {Physical Review B}\ }\textbf {\bibinfo {volume} {99}},\ \bibinfo {pages}
	  {184510} (\bibinfo {year} {2019})}\BibitemShut {NoStop}%
	\bibitem [{\citenamefont {Klett}\ \emph {et~al.}(2020)\citenamefont {Klett},
	  \citenamefont {Wentzell}, \citenamefont {Sch{\"a}fer}, \citenamefont
	  {Simkovic~IV}, \citenamefont {Parcollet}, \citenamefont {Andergassen},\ and\
	  \citenamefont {Hansmann}}]{klett2020real}%
	  \BibitemOpen
	  \bibfield  {author} {\bibinfo {author} {\bibfnamefont {Marcel}\ \bibnamefont
	  {Klett}}, \bibinfo {author} {\bibfnamefont {Nils}\ \bibnamefont {Wentzell}},
	  \bibinfo {author} {\bibfnamefont {Thomas}\ \bibnamefont {Sch{\"a}fer}},
	  \bibinfo {author} {\bibfnamefont {Fedor}\ \bibnamefont {Simkovic~IV}},
	  \bibinfo {author} {\bibfnamefont {Olivier}\ \bibnamefont {Parcollet}},
	  \bibinfo {author} {\bibfnamefont {Sabine}\ \bibnamefont {Andergassen}}, \
	  and\ \bibinfo {author} {\bibfnamefont {Philipp}\ \bibnamefont {Hansmann}},\
	  }\bibfield  {title} {\enquote {\bibinfo {title} {Real-space cluster dynamical
	  mean-field theory: Center-focused extrapolation on the one-and two
	  particle-levels},}\ }\href@noop {} {\bibfield  {journal} {\bibinfo  {journal}
	  {Physical Review Research}\ }\textbf {\bibinfo {volume} {2}},\ \bibinfo
	  {pages} {033476} (\bibinfo {year} {2020})}\BibitemShut {NoStop}%
	\bibitem [{\citenamefont {Verret}\ \emph {et~al.}(2019)\citenamefont {Verret},
	  \citenamefont {Roy}, \citenamefont {Foley}, \citenamefont {Charlebois},
	  \citenamefont {S{\'e}n{\'e}chal},\ and\ \citenamefont
	  {Tremblay}}]{verret2019intrinsic}%
	  \BibitemOpen
	  \bibfield  {author} {\bibinfo {author} {\bibfnamefont {S}~\bibnamefont
	  {Verret}}, \bibinfo {author} {\bibfnamefont {J}~\bibnamefont {Roy}}, \bibinfo
	  {author} {\bibfnamefont {A}~\bibnamefont {Foley}}, \bibinfo {author}
	  {\bibfnamefont {M}~\bibnamefont {Charlebois}}, \bibinfo {author}
	  {\bibfnamefont {D}~\bibnamefont {S{\'e}n{\'e}chal}}, \ and\ \bibinfo {author}
	  {\bibfnamefont {A-MS}\ \bibnamefont {Tremblay}},\ }\bibfield  {title}
	  {\enquote {\bibinfo {title} {Intrinsic cluster-shaped density waves in
	  cellular dynamical mean-field theory},}\ }\href@noop {} {\bibfield  {journal}
	  {\bibinfo  {journal} {Physical Review B}\ }\textbf {\bibinfo {volume}
	  {100}},\ \bibinfo {pages} {224520} (\bibinfo {year} {2019})}\BibitemShut
	  {NoStop}%
	\bibitem [{\citenamefont {Fratino}\ \emph {et~al.}(2016)\citenamefont
	  {Fratino}, \citenamefont {S{\'e}mon}, \citenamefont {Sordi},\ and\
	  \citenamefont {Tremblay}}]{fratino2016organizing}%
	  \BibitemOpen
	  \bibfield  {author} {\bibinfo {author} {\bibfnamefont {L}~\bibnamefont
	  {Fratino}}, \bibinfo {author} {\bibfnamefont {P}~\bibnamefont {S{\'e}mon}},
	  \bibinfo {author} {\bibfnamefont {Giovanni}\ \bibnamefont {Sordi}}, \ and\
	  \bibinfo {author} {\bibfnamefont {A-MS}\ \bibnamefont {Tremblay}},\
	  }\bibfield  {title} {\enquote {\bibinfo {title} {An organizing principle for
	  two-dimensional strongly correlated superconductivity},}\ }\href@noop {}
	  {\bibfield  {journal} {\bibinfo  {journal} {Scientific reports}\ }\textbf
	  {\bibinfo {volume} {6}},\ \bibinfo {pages} {1--6} (\bibinfo {year}
	  {2016})}\BibitemShut {NoStop}%
	\end{thebibliography}

\end{document}